\documentclass[twocolumn,showpacs,eqsecnum,amsmath,amssymb,prd]{revtex4}

\usepackage{graphicx}
\usepackage{dcolumn}
\usepackage{bm}
\usepackage{psfrag}

\usepackage{amsmath,epsfig,amsfonts,color,lscape}

\newcommand{\1}{\mbox{1}\hspace{-0.25em}\mbox{l}}

\begin{document}

\preprint{}

\title{
Index theorem and Majorana zero modes along a non-Abelian vortex\\
in a color superconductor
}

\author{Takanori Fujiwara,$^1$ Takahiro  Fukui,$^1$ Muneto Nitta,$^2$ and
Shigehiro Yasui$^3$}
\affiliation{$^1$Department of Physics, Ibaraki University, Mito
310-8512, Japan\\
$^2$Department of Physics, and Research and Education Center for Natural Sciences, 
Keio University, 4-1-1 Hiyoshi, Yokohama, Kanagawa 223-8521, Japan\\
$^3$KEK Theory Center, Institute of Particle and Nuclear Studies, 
High Energy Accelerator Research Organization (KEK), 1-1 Oho,
Tsukuba, Ibaraki, 305-0801, Japan}

\date{\today}

\begin{abstract}
Color superconductivity  in high density QCD exhibits
the color-flavor locked (CFL) phase.
To explore zero modes in the CFL phase
in the presence of a non-Abelian vortex
with an  SU(2) symmetry in the vortex core, 
we apply the index theorem to the Bogoliubov-de Gennes (BdG) Hamiltonian.
From the calculation of the topological index,  
we find that triplet, doublet and singlet sectors of SU(2)   
have certain number of chiral Majorana zero modes 
in the limit of vanishing chemical potential.
We also solve the BdG equation by the use of the series expansion to show that
the number of zero modes and their chirality match the result of the index theorem.   
From particle-hole symmetry of the BdG Hamiltonian,
we conclude that if and only if the index of a given sector is odd, 
one 
zero mode survives generically for a finite chemical potential.
We argue that this result should hold nonperturbatively even in the high density limit.
\end{abstract}

\maketitle

\section{Introduction}

Zero modes around a vortex in a superconductor have been attracting much current interest,
providing us with intriguing notion of 
the Majorana fermion. 
Vortices in a $p_x+ip_y$ superconductor
yield zero-energy Majorana bound states \cite{Volovik:99} 
which obey non-Abelian statistics \cite{ReadGreen:00,Kitaev:00,Ivanov:01,SOM04,SNT06,TSL07,GurRad07}.
Generically, these degenerate states are expected to be quite useful in 
constructing a fault-tolerant quantum computer \cite{Kit06,Kit07}, and 
the idea for topological quantum computation utilizing
non-Abelian anyons on a topological state of matter
has been developed
\cite{Nayak:2008zza}.
Majorana states have been predicted in many other systems such as a 
surface state of a topological insulator
with the proximity effect of an $s$-wave superconductor \cite{Fu:2008fk}
or of a ferromagnetic insulator \cite{TYN09},
an $s$-wave superfluid of ultra-cold atoms \cite{STF09}, 
the superfluid $^3$He B-phase \cite{Vol09}, etc
\cite{QHRZ09,BerHur09,SLTS10,Lee09,Ali10,Her10,LTYSN10,Her09b}.
Not only in condensed matter physics but also 
in high density QCD with color superconductivity \cite{Alford:1998mk,Alford:2007xm},
Majorana zero modes along an Abelian vortex \cite{Nishida:10} and 
along a non-Abelian vortex \cite{YasuiIN:10} in the CFL 
phase have been discussed,
and new non-Abelian statistics has been derived \cite{Yasui:2010yh}.
Fermionic zero modes in the presence of topological background
have also been explored in the standard model of 
the electro-weak interaction 
\cite{Earnshaw:1994jj,Moreno:1994bk,Garriga:1994wb,Stoj00,SSV01,SSV02}. 
Recently, a classification scheme for zero modes 
associated with topological defects has been proposed \cite{TeoKane:10,TeoKane:10b}, 
which is a generalization of the topological classification for 
insulators and superconductors \cite{SchnyderRFL:08,Kitaev:08}. 

Various methods of counting zero modes in such systems
have been proposed
\cite{Callias:78,Weinberg:81,Kit00,Kit06,TSS10,TeoKan10,FukuiFujiwara:10,HCM10,SNCM10,SRCM09,SLTS09}.
Among them, application \cite{FukuiFujiwara:10} of the index theorem 
\cite{Callias:78,Weinberg:81}
is unique in that it does not resort to any approximations.
The index theorem has a long history and 
has elucidated, for example, topological characteristics of anomalies in gauge theories
\cite{EGH80}.
The index theorem applied to the present problem 
is a variant, dealing with a fermion coupled to a condensate
with a nontrivial topological defect.
This rigorous theorem claims 
equivalence between following two quantities: 
One is associated with the number of zero modes
which reflects the analytical property of a differential equation.
This can be obtained if one indeed solves the eigenvalue equation 
\cite{JacReb76,JacRos81}.  
The other is associated  with a topological invariant of 
the order parameter with a defect. 
The index theorem thus relates the zero modes with the topological configuration of a defect. 
It has been applied to a fermion describing a
topological superconductor in 3D \cite{Fukui10}, 
and generalized to a $\mathbb{Z}_2$ index theorem for
a fermion without chiral symmetry \cite{FukuiFujiwara10b}.

In this paper, we apply the index theorem to the CFL phase 
of a color superconductor with
a non-Abelian vortex 
$\Delta = {\rm diag}(\Delta_Q,\Delta_Q,\Delta_q)$
with an  SU(2) symmetry in the vortex core, 
where $\Delta_q$ and $\Delta_Q$ have winding 
$q$ and $Q$, respectively.
We find that triplet, doublet and singlet sectors of SU(2)   
have $q$, $Q$, and $2Q$ fermionic zero modes, respectively, 
propagating along a vortex line 
in the limit of vanishing chemical potential.
These modes can be regarded as one-dimensional chiral Majorana fermions 
due to particle-hole symmetry.  
We obtain these results from the calculation of the topological index as well as
from the analysis of the normalizability of 
the zero mode wave functions,
and find complete agreement between them.
We argue that 
when the chemical potential is switched on, 
no (or more precisely, even) zero modes are expected 
in the case of an even index, 
whereas one (or more precisely, odd) zero mode survives in the case of an odd index.
This holds in each sector of SU(2), and
is due to spectral symmetry 
between positive- and negative-energy states 
of the BdG Hamiltonian.
Therefore, it should be
extrapolated even into very high density limit nonperturbatively.

This paper is organized as follows.
In the next section \ref{s:EffHam}, we derive effective Hamiltonians for triplet,
doublet, and singlet states of SU(2) from the whole BdG Hamiltonian.
In Sec. \ref{s:TopInd}, we give a brief review of the index theorem relevant to 
the present problem, and calculate the topological index for all sectors of SU(2). 
In Sec. \ref{s:AnaInd}, we switch to solving the BdG equation, and 
count the zero modes and their chiralities directly, 
from which we know the analytical index.
We show that these two indices indeed coincide. 
In the final section \ref{s:Sum}, summary and discussions are given.
In Appendix \ref{s:singlet}, we show that a singlet zero mode found in 
\cite{YasuiIN:10} together with a new solution diverges in a short distance 
and is non-normalizable 
although it was shown to converge in a large distance.
In Appendix \ref{s:App}, we give a brief review  
of the way to solve 
the differential equations of zero modes in a series expansion.

\section{Effective Hamiltonian}
\label{s:EffHam} 

Under a non-Abelian vortex with an SU(2) symmetry, 
all eigenstates in the CFL phase
are decomposed into multiplets of SU(2). 
Therefore, for later convenience, we reduce
BdG Hamiltonian into smaller ones which act on each multiplet.

\subsection{BdG equation and vortices}
\label{s:BdG}

We start with a generic Hamiltonian with a given pairing gap of superconductivity,  
\begin{alignat}1
H=&\int d^3x
\Big[\psi_i^{\alpha\dagger}\left(-i\gamma^0\bm\gamma\cdot\nabla-\mu\right)\psi_i^\alpha
\nonumber\\
&\quad+\Delta_{ij}^{\alpha\beta}(\psi_i^{\alpha \dagger}\gamma_5{\cal C}^\dagger\psi_j^{\beta*})
+\Delta_{ij}^{\alpha\beta*}(\psi_j^{\beta T}{\cal C}\gamma_5\psi_i^{\alpha})\Big]
\nonumber\\
=&\int d^3x\hat\psi_i^{\alpha\dagger}(\hat{\cal H}_{\rm tot})_{ij}^{\alpha\beta}\hat\psi_j^\beta ,
\end{alignat}
where $\alpha$, $\beta$ and $i$, $j$ are color and flavor indices, respectively,
${\cal C}=i\gamma^2\gamma^0$ 
is the charge conjugation matrix, and 
$\Delta_{ij}^{\alpha\beta}$ is a generic gap function with 
$\Delta_{ij}^{\alpha\beta}=\Delta_{ji}^{\beta\alpha}$ due to Fermi statistics.
In the last line, we have introduced the Nambu representation
\begin{eqnarray}
\hat\psi_i^\alpha=\left(\psi_i^\alpha,(\psi_i^\alpha)^c\right)^T,
\end{eqnarray}
where 
$(\psi_i^\alpha)^c\equiv{\cal C}\bar\psi^T= i\gamma^2\psi_i^{\alpha*}$ is the charge conjugation of
$\psi_i^\alpha$, and the total Hamiltonian density $\hat{\cal H}_{\rm tot}$ is defined by
\begin{alignat}1
(\hat{\cal H}_{\rm tot})_{ij}^{\alpha\beta}&
=\hat{\cal H}_0\delta_{ij}\delta^{\alpha\beta}+\hat\Delta_{ij}^{\alpha\beta} ,
\end{alignat}
with
\begin{alignat}1
\hat{\cal H}_0&=\left(
\begin{array}{cc}
-i\gamma^0\bm\gamma\cdot\nabla-\mu&\\
&-i\gamma^0\bm\gamma\cdot\nabla+\mu
\end{array}\right) 
\nonumber\\
&\equiv\left(
\begin{array}{cc}
{\cal H}_0-\mu&\\
&{\cal H}_0+\mu
\end{array}\right) ,
\label{Ham0}\\ 
\hat\Delta_{ij}^{\alpha\beta}&=\left(
\begin{array}{cc}
&\Delta_{ij}^{\alpha\beta}\gamma_5\gamma^0\\
-\Delta_{ij}^{\alpha\beta *}\gamma_5\gamma^0&
\end{array}\right) .
\nonumber
\end{alignat}
In what follows, the hat means the Nambu representation with particle-hole symmetry,
where $\psi_i^\alpha$ and $(\psi_i^\alpha)^c$ are referred to as a particle and a hole, 
respectively.
This Hamiltonian density leads us to the BdG eigenvalue equation,
\begin{alignat}1
\hat{\cal H}_{\rm tot}\hat\Psi=E\hat\Psi .
\label{GenBogDegEqu}
\end{alignat}
Counting the zero modes of this equation is the main interest of the present paper.
Here, the zero mode means a massless mode propagating along a vortex which 
will be specified momentarily.

Now we take three flavors ($i,j=$) $u,d,$ and $s$, 
and assume the following gap function
in the CFL phase
\begin{alignat}1
\Delta_{ij}^{\alpha\beta}=\epsilon^{\alpha\beta\gamma}\epsilon_{ijk}\Delta_\gamma^k ,
\end{alignat}
which belongs to $\bar3$ representation with respect to SU(3)$_{\rm C}$ 
as well as SU(3)$_{\rm F}$.
Under the action of 
$(e^{i\theta},U_{\rm C},U_{\rm F}) \in 
{\rm U(1)}_{\rm B} \times {\rm SU(3)}_{\rm C}\times {\rm SU(3)}_{\rm F}$,
the gap function transforms as
\begin{eqnarray}
&& \Delta_\gamma^k 
 \to e^{i\theta}(U_{\rm C})_\gamma^\alpha 
     \Delta_\alpha^i (U_{\rm F}^T)_i^k  .
\end{eqnarray}
The gap function takes a form of 
\begin{alignat}1
\Delta_\gamma^k = \Delta_{\rm CFL}{\bf 1} 
\label{CFLGro}
\end{alignat}
in the ground state, where $\Delta_{\rm CFL}$ is a constant.
In this state, the so-called color-flavor locking 
symmetry ${\rm SU}(3)_{\rm C+F}$, which acts on the gap as 
\begin{eqnarray}
\Delta_\gamma^k \to (U \Delta U^\dagger)_\gamma^k
\label{eq:CFL}
\end{eqnarray}
with $U \in {\rm SU}(3)$, is preserved.

We consider a vortex state in the CFL phase in this paper.
The minimal winding non-Abelian vortex, 
denoted by $M_1$ in \cite{Balachandran:2005ev}, is given by  
\begin{eqnarray}
\Delta_\gamma^k
=\mbox{diag}\left(\Delta_0, \Delta_0,\Delta_1\right) ,
\end{eqnarray}
where $\Delta_q(r,\theta)=|\Delta_q(r)|e^{iq\theta}$ stands for a gap function
with a $q$-vorticity which obeys 
the boundary conditions 
$|\Delta_0(\infty)|=|\Delta_1(\infty)|=\Delta_{\rm CFL}$ and
$|\Delta_0(0)|' = |\Delta_1(0)|=0$.
This vortex carries a 1/3 quantized circulation of $U(1)_{\rm B}$ 
and a color magnetic flux is confined inside its core 
which is neglected in the BdG equation.
The CFL symmetry ${\rm SU}(3)_{\rm C+F}$ in 
Eq.~(\ref{eq:CFL}) is spontaneously broken 
into ${\rm SU}(2)_{\rm C+F}\times {\rm U}(1)_{\rm C+F}$ around the core of the vortex. 
Consequently, Nambu-Goldstone bosonic zero modes 
${\mathbb C}P^2 \simeq {\rm SU}(3)_{\rm C+F}/[{\rm SU}(2)_{\rm C+F}
\times {\rm U}(1)_{\rm C+F}]$  
appear inside the core of the vortex
\footnote{
However, these modes are gapped if one takes into account
the non-perturbative effect in 
the vortex world-sheet theory \cite{Gorsky:2011hd}.
}
\cite{Nakano:2007dr,Eto:2009kg,Eto:2009tr,Hirono:2010gq}.
Properties of this vortex has been extensively studied in 
the Ginzburg-Landau model
\cite{Balachandran:2005ev,Nakano:2007dr,Eto:2009kg,Eto:2009tr,
Hirono:2010gq,Gorsky:2011hd}, 
and in the BdG equation \cite{YasuiIN:10,Yasui:2010yh}.
On the other hand, a non-Abelian vortex, 
denoted by $M_2$ \cite{Balachandran:2005ev},
\begin{eqnarray}
\Delta_\gamma^k
=\mbox{diag}(\Delta_1,\Delta_1,\Delta_0) ,\label{eq:M2}
\end{eqnarray}
with the same boundary conditions on the diagonal elements $\Delta_q$ as in $M_1$,
preserves the same symmetry 
${\rm SU}(2)_{\rm C+F}\times {\rm U}(1)_{\rm C+F}$ in the vortex core 
\footnote{
In Ref.~\cite{Balachandran:2005ev} 
anti-vortex of (\ref{eq:M2}) was referred to as $M_2$ 
because it has the same color magnetic flux with $M_1$,
whereas $M_2$ here in Eq.~(2.11)
has the flux of the same magnitude with $M_1$ with a minus sign.
}.
This vortex carries a 2/3 quantized circulation of $U(1)_{\rm B}$.

More generally, in this paper, 
we consider a bound state of 
$q$ $M_1$ and $Q$ $M_2$ non-Abelian vortices, 
for which the gap function takes the form of
\begin{alignat}1
\Delta_\gamma^k=\mbox{diag}(\Delta_Q,\Delta_Q,\Delta_q) .
\label{NonAbeVor}
\end{alignat}
The case of $q=1$ and $Q=0$ is the non-Abelian $M_1$ vortex above, 
while $q=0$ and $Q=1$ corresponds to the non-Abelian $M_2$ vortex. 
When $q\neq Q$,
this vortex breaks SU(3)$_{\rm C+F}$ to SU(2)$_{\rm C+F}$$\times$ U(1)$_{\rm C+F}$. 
Therefore, the BdG eigenstates in (\ref{GenBogDegEqu}) 
are classified as the SU(2) multiplets.
To derive an effective Hamiltonian in each multiplet is the main task in this section.

For a generic gap function in Eq. (\ref{NonAbeVor}), 
we adopt the basis 
$\hat\psi=(\hat u_r,\hat d_g,\hat s_b,\hat d_r,\hat u_g,\hat s_r,\hat u_b,\hat s_g,\hat d_b)^T$
to write down the BdG equation in Eq.~(\ref{GenBogDegEqu})
such that \cite{Alford:1999pa,Sadzikowski:2002in}
\begin{widetext}
\begin{alignat}1
\left(
\begin{array}{ccccccccc}
\hat{\cal H}_0 & \hat\Delta_q  &\hat\Delta_Q    &&&&&&\\
\hat\Delta_q   &\hat{\cal H}_0 & \hat\Delta_Q   &&&&&&\\
\hat\Delta_Q   &\hat\Delta_Q   &\hat{\cal H}_0  &&&&&&\\
&&& \hat{\cal H}_0 &-\hat\Delta_q  &&&&\\
&&&-\hat\Delta_q   &\hat{\cal H}_0 &&&&\\
&&&&& \hat{\cal H}_0 &-\hat\Delta_Q  &&\\
&&&&&-\hat\Delta_Q   &\hat{\cal H}_0 &&\\
&&&&&&& \hat{\cal H}_0 &-\hat\Delta_Q  \\
&&&&&&&-\hat\Delta_Q   &\hat{\cal H}_0 \\
\end{array}
\right)
\left(
\begin{array}{c}
\hat u_r\\
\hat d_g\\
\hat s_b\\
\hat d_r\\
\hat u_g\\
\hat s_r\\
\hat u_b\\
\hat s_g\\
\hat d_b
\end{array}
\right)
=E
\left(
\begin{array}{c}
\hat u_r\\
\hat d_g\\
\hat s_b\\
\hat d_r\\
\hat u_g\\
\hat s_r\\
\hat u_b\\
\hat s_g\\
\hat d_b
\end{array}
\right),
\label{HamTot}
\end{alignat}
\end{widetext}
where $\hat\Delta_q$ is defined by 
\begin{alignat}1
&\hat\Delta_q=
\left(
\begin{array}{cc}
&\Delta_q\gamma_5\gamma^0\\
-\Delta_q^*\gamma_5\gamma^0&\\
\end{array}
\right) ,
\label{GapCFL}
\end{alignat}
and $\hat\Delta_Q$ similarly.

Let us now derive an effective Hamiltonian which acts on each multiplet 
of SU(2)$_{\rm C+F}$.
To this end, it is convenient to write the particle states as a $3\times3$ matrix via
$\psi_i^\alpha=(\psi)_{i\alpha}$,
\begin{alignat}1
\left(
\begin{array}{ccc}
u_r & u_g & u_b\\
d_r & d_g & d_b\\
s_r & s_g & s_b
\end{array}
\right) =\sum_{a=1}^9\Psi_aT_a ,
\label{ParMatSta}
\end{alignat}
where 
$T_a$ $(a=1,\cdots,8)$ is the generator of SU(3) normalized by ${\rm tr}\,T_aT_b=\delta_{ab}/2$
defined through the Gell-Mann matrices as $T_a=\lambda_a/2$ and $T_9=(1/\sqrt{6}\cdot\bm1)$.
This equation implies that the particle states of the CFL phase
can be regarded,
if they are in the ground state specified by Eq. (\ref{CFLGro}),
as the adjoint representation of SU(3)$_{\rm C+F}$,
$\sum_{a=1}^8\Psi_aT_a$ and an invariant state of a U(1) subalgebra $\Psi_9T_9$.
The non-Abelian vortex in Eq.~(\ref{NonAbeVor}) tells that 
we can reduce  this representation into those which are irreducible under 
SU(2)$_{\rm C+F}$ generated by the subalgebra $(T_1,T_2,T_3)$.
Under such subalgebra,
$(T_1,T_2,T_3)$ themselves transform as the adjoint (spin-1) representation,
$(T_4+iT_5,T_6+iT_7)$ and
$(T_6-iT_7,T_4-iT_5)$ transform as the spin-1/2 representation,
and $T_8$ and $T_9$ transform as spin-0 representation.
Therefore, the states (\ref{ParMatSta}) are decomposed into
one triplet, two doublets, and two singlets such that \cite{YasuiIN:10}
\begin{alignat}1
&
\Psi_{\rm t}= u_g (T_1+iT_2)+ (u_r - d_g)T_3+ d_r(T_1-iT_2) ,
\nonumber\\
&
\Psi_{\rm d1}= u_b(T_4+iT_5)+ d_b (T_6+iT_7),
\nonumber\\
&
\Psi_{\rm d2}= s_g (T_6-iT_7)+ s_r(T_4-iT_5),
\nonumber\\
&
\Psi_{\rm s1}=\frac{u_r+d_g-2 s_b}{\sqrt{3}}T_8,
\nonumber\\
&
\Psi_{\rm s2}=2\frac{u_r+ d_g+ s_b}{\sqrt{6}}T_9 .
\label{ParSU2Mul}
\end{alignat}
Likewise, the hole states can be written as
\begin{alignat}1
\left(
\begin{array}{ccc}
u_r^c & u_g^c & u_b^c\\
d_r^c & d_g^c & d_b^c\\
s_r^c & s_g^c & s_b^c
\end{array}
\right) =\sum_{a=1}^9\Psi_a^cT_a^* ,
\label{HolMatSta}
\end{alignat}
since they are complex conjugate of the particle states,
$\psi^c=i\gamma^2\psi^*$. Therefore, corresponding SU(2)$_{\rm C+F}$ multiplets are
\begin{alignat}1
&
\Psi_{\rm t}^c= d_r^c (T_1+iT_2)+ (u_r^c - d_g^c)T_3+ u_g^c(T_1-iT_2) ,
\nonumber\\
&
\Psi_{\rm d1}^c= d_b^c(T_4+iT_5)+ u_b^c (T_6+iT_7),
\nonumber\\
&
\Psi_{\rm d2}^c= s_r^c (T_6-iT_7)+ s_g^c(T_4-iT_5),
\nonumber\\
&
\Psi_{\rm s1}^c=\frac{u_r^c+d_g^c-2 s_b^c}{\sqrt{3}}T_8,
\nonumber\\
&
\Psi_{\rm s2}^c=2\frac{u_r^c+ d_g^c+ s_b^c}{\sqrt{6}}T_9 .
\label{HolSU2Mul}
\end{alignat}
It is now easy to see that 
the BdG equation in (\ref{HamTot}) is decoupled into small pieces acting on 
the above multiplets, as will be done below.
Among them, our primary concern is the singlet states.

\subsection{SU(2) doublet sector}

The two doublet states, whose spaces are spanned by 
$\hat u_b,\hat d_b,\hat s_g$, and $\hat s_r$,
are coupled only to the gap $\Delta_Q$.
Equation (\ref{HamTot}) shows that these are decoupled into $(\hat s_r, \hat u_b)$ and
$(\hat s_g,\hat d_b)$. 
Considering $\hat{\cal H}_0$ in Eq.~(\ref{Ham0}) and $\hat \Delta_Q$ in Eq.~(\ref{GapCFL}),
we see that among the states $(\hat s_r,\hat u_b)$, for example, 
$(s_r,u_b^c)$ and $(u_b,s_r^c)$ are decoupled, and $(\hat s_g,\hat d_b)$ likewise.
This means that 
the Hamiltonian for $(\hat s_r,\hat u_b)$ and $(\hat s_g, \hat d_b)$ 
is block-diagonal, and BdG equation is decomposed into 
\begin{alignat}1
&
\left(
\begin{array}{cc}
{\cal H}_0-\mu&-\Delta_Q\gamma_5\gamma^0\\
\Delta_Q^*\gamma_5\gamma^0&{\cal H}_0+\mu
\end{array}
\right)
\left(
\begin{array}{c}
\varphi \\ \eta^c
\end{array}
\right)
=E
\left(
\begin{array}{c}
\varphi \\ \eta^c
\end{array}
\right) ,
\label{NaiDouHam1}
\end{alignat}
where 
\begin{alignat}1
(\varphi,\eta^c)\equiv(s_r,u_b^c), (u_b,s_r^c), (s_g,d_b^c), (d_b,s_g^c).  
\label{NaiDouHam2}
\end{alignat}
It should be noted that the Hamiltonian in Eq. (\ref{NaiDouHam1}) can be 
simply given by 
\begin{alignat}1
\hat{\cal H}_0-\hat\Delta_Q .
\label{NaiEffHamDou}
\end{alignat}
Thus, this can be regarded as the effective Hamiltonian in the doublet states.
We here mention that from Eqs. (\ref{ParSU2Mul}) and (\ref{HolSU2Mul}), 
the SU(2) quantum number $T_3$ reads $T_3=-1/2$, $1/2$, $1/2$, and $-1/2$, respectively,
for the paired states $(s_r,u_b^c)$, $(u_b,s_r^c)$, $(s_g,d_b^c)$, and $(d_b,s_g^c)$.
This implies that the initial particle and hole paired states, e.g., $(s_r,s_r^c)$, switch 
their partners to $(s_r,u_b^c)$ and so on in order 
to be the simultaneous eigenstates of $T_3$ as well as the Hamiltonian.
In this representation, the new paired states
are related by
\begin{alignat}1
\left(
\begin{array}{c}
\varphi\\ \eta^c
\end{array}
\right) =C\left(
\begin{array}{c}
\eta \\ \varphi^c 
\end{array}
\right) ,
\end{alignat}
where anti-unitary particle-hole transformation $C$ is defined by 
\begin{alignat}1
C=\left(\begin{array}{cc}&i\gamma^2\\ i\gamma^2&\end{array}\right)K,\quad
C^2=1 ,
\label{ParHolSymDou}
\end{alignat}
with complex conjugation $K$.
Correspondingly, 
the effective Hamiltonian (\ref{NaiEffHamDou}) has indeed particle-hole symmetry, 
\begin{alignat}1
& C(\hat{\cal H}_0-\hat\Delta_Q)C^{-1} =-(\hat{\cal H}_0-\hat\Delta_Q),
\end{alignat}
as it should do.
Although 
the BdG equation in the doublet states is described by single equation (\ref{NaiDouHam1}),
the above particle-hole symmetry implies that if $(\varphi,\eta^c)$ is regarded as 
particle states with an energy $E$, 
corresponding hole states are given by $(\eta,\varphi^c)$ with energy $-E$.
The Hamiltonian (\ref{NaiEffHamDou}), which appears 
in Eq. (\ref{NaiDouHam1}),
can thus be regarded as an effective Hamiltonian involved in the SU(2) doublets
with correct SU(2) quantum numbers $(\sum_{a=1,2,3}T_a^2,T_3)$ and 
with correct particle-hole symmetry described by $C$.

This Hamiltonian can be further decomposed into right- and left-handed modes,
i.e., the eigenstates of $\gamma_5$ with eigenvalues $\gamma_5=\pm1$.
Let $(\varphi,\eta^c)$ be a paired SU(2) doublet state defined in Eq. (\ref{NaiDouHam2}),
and let $\varphi=(\varphi_+,\varphi_-)$ and $\eta^c=(\eta^c_-,\eta^c_+)$ be
the decomposition into right and left modes, where $+$ and $-$ mean 
the right and left modes, respectively. 
We use the following $\gamma$ matrices in this paper,
\begin{alignat}1
\gamma^0=
\left(
\begin{array}{cc}
&1\\1&
\end{array}
\right),
~
\bm\gamma=
\left(
\begin{array}{cc}
&-\bm\sigma\\\bm\sigma&
\end{array}
\right),
~
\gamma_5=
\left(
\begin{array}{cc}
1&\\&-1
\end{array}
\right).
\end{alignat}
Then, the BdG equation (\ref{NaiDouHam1}) can be decomposed into
\begin{alignat}1
\left(
\begin{array}{cc}
-i\bm\sigma\cdot\nabla-\mu&-\Delta_Q\\
-\Delta_Q^* & i\bm\sigma\cdot\nabla+\mu 
\end{array}
\right)
\left(\begin{array}{c}\varphi_+\\\eta^c_+\end{array}\right)
=E\left(\begin{array}{c}\varphi_+\\\eta^c_+\end{array}\right) ,
\end{alignat}
and 
\begin{alignat}1
\left(
\begin{array}{cc}
i\bm\sigma\cdot\nabla-\mu&\Delta_Q\\
\Delta_Q^* & -i\bm\sigma\cdot\nabla+\mu 
\end{array}
\right)
\left(\begin{array}{c}\varphi_-\\\eta^c_-\end{array}\right)
=E\left(\begin{array}{c}\varphi_-\\\eta^c_-\end{array}\right) .
\end{alignat}
These equations define effective Hamiltonians $\hat{\cal H}_{\rm d}^\pm$
which act on SU(2) doublets  with $\gamma_5=\pm1$ chirality,
\begin{alignat}1
\hat{\cal H}_0-\hat\Delta_Q&\rightarrow
\left(\begin{array}{cc}\hat{\cal H}_0^+-\hat\Delta_Q^+&\\&
 \hat{\cal H}_0^--\hat\Delta_Q^-\end{array}\right)
\nonumber\\
&\equiv
\left(\begin{array}{cc}\hat{\cal H}_{\rm d}^+&\\&
 \hat{\cal H}_{\rm d}^-\end{array}\right),
\label{ChiDecHam}
\end{alignat}
where 
\begin{alignat}1
&\hat{\cal H}_0^\pm=\pm
\left(
\begin{array}{cc}
-i\bm\sigma\cdot\nabla&\\
&i\bm\sigma\cdot\nabla
\end{array}
\right)
-\mu\left(
\begin{array}{cc}
1&\\
&-1
\end{array}
\right) ,
\nonumber\\
&\hat\Delta_Q^\pm=\pm
\left(
\begin{array}{cc}
&\Delta_Q\\
\Delta_Q^*&
\end{array}
\right) .
\label{LefRig}
\end{alignat}

When we calculate the index in the next section, 
it is convenient to define new $\Gamma$ matrices such that
\begin{alignat}1
&\Gamma^j=\sigma^j\otimes\sigma^3,\quad (j=1,2,3)
\nonumber\\
&\Gamma^4=1\otimes\sigma^1 ,
\nonumber\\
&\Gamma^5=1\otimes\sigma^2 ,
\label{DouGamMat}
\end{alignat}
which satisfy the relation $\{\Gamma^\mu,\Gamma^\nu\}=2\delta^{\mu\nu}$.
Then, the Hamiltonians in the above can be written as
\begin{alignat}1
\hat{\cal H}_{\rm d}^\pm=\pm\hat{\cal H}_{\rm d}+i\mu\Gamma^4\Gamma^5,
\label{DouChiHam}
\end{alignat}
where 
\begin{alignat}1
\hat{\cal H}_{\rm d}&=
\left(
\begin{array}{cc}
-i\bm\sigma\cdot\nabla&-\Delta_Q\\-\Delta_Q^*&i\bm\sigma\cdot\nabla
\end{array}
\right)
\nonumber\\
&=-i\Gamma^j\partial_j+\Gamma^a\phi_a,
\label{DouChiHamCom}
\end{alignat}
with $j=1,2,3$, $a=4,5$, and 
\begin{alignat}1
\bm\phi&\equiv(\phi_4,\phi_5)
\nonumber\\
&= (-{\rm Re}\,\Delta_Q,{\rm Im}\,\Delta_Q) .
\label{DouOrdPar}
\end{alignat}
The Hamiltonian $\hat{\cal H}_{\rm d}^\pm$ in Eq. (\ref{DouChiHam})
is the final expression of the 
effective Hamiltonian in the doublet states, and $\hat{\cal H}_{\rm d}$
in Eq. (\ref{DouChiHamCom}) plays a crucial role in the index theorem
in the next section.

This Hamiltonian belongs to the generic universality class of superconductors, 
class D \cite{SchnyderRFL:08,Kitaev:08}, since it transforms 
\begin{alignat}1
&
\widetilde C\hat{\cal H}_{\rm d}^\pm \widetilde C^{-1}=-\hat{\cal H}_{\rm d}^\pm,
\label{ParHolTri}
\end{alignat}
under a reduced particle-hole transformation $\widetilde C$  defined by
\begin{alignat}1
\widetilde C=i\Gamma^2\Gamma^4K,\quad \widetilde C^2=1,
\end{alignat}
which is associated with Eq. (\ref{ParHolSymDou}),
acting on the spaces of right or left modes separately.

\subsection{SU(2) triplet sector}

The triplet states couple only to the gap $\Delta_q$, and when $q=1$,
the explicit solutions of the zero modes have already been obtained in \cite{YasuiIN:10}. 
Equations (\ref{ParSU2Mul}) and (\ref{HolSU2Mul}) tell that 
three states $\hat u_g$, $\hat d_r$, and $\hat u_r-\hat d_g$ are involved in this sector.
As to the $T_3=\pm1$ states, $\hat u_g$ and $\hat d_r$, it follows from Eq. (\ref{HamTot})
that they obey the same BdG equation  
as in the doublet sector if the gap function is replaced by $\hat\Delta_q$, and 
therefore, an effective Hamiltonian for paired state $(d_r,u_g^c)$ and $(u_g,d_r^c)$ 
is given by
\begin{alignat}1
\hat{\cal H}_0-\hat\Delta_q,
\label{NaiEffHamTri}
\end{alignat}
in the same sense as in Eq.~(\ref{NaiEffHamDou}). 
As to the $T_3=0$ state, $\hat u_r-\hat d_g$, 
it is easy to derive, from the BdG equation in Eq.~(\ref{HamTot}),
the following equation;
\begin{alignat}1
(\hat{\cal H}_0-\hat\Delta_q)(\hat u_r-\hat d_g)=E(\hat u_r-\hat d_g),
\end{alignat}
whose Hamiltonian is just the same as the one in Eq.~(\ref{NaiEffHamTri}).
Therefore, the Hamiltonian (\ref{NaiEffHamTri}) can be regarded as an effective
Hamiltonian in the triplet sector. 
Chiral decomposition is also carried out in the same way as in the previous 
subsection, and the final Hamiltonian, which is referred to as $\hat{\cal H}_{\rm t}^\pm$, 
is given by Eqs. (\ref{DouChiHam}) and (\ref{DouChiHamCom}) with 
a different order parameter
\begin{alignat}1
\bm\phi&\equiv(\phi_4,\phi_5)
\nonumber\\
&= (-{\rm Re}\,\Delta_q,{\rm Im}\,\Delta_q) .
\label{TriOrdPar}
\end{alignat}
This Hamiltonian also belongs to class D.

\subsection{SU(2) singlet sector}

While the doublet and triplet sectors have rather simple Hamiltonian basically 
equivalent to single flavor case \cite{YasuiIN:10}, the singlet sectors couple to 
gaps with different winding, which could yield nontrivial zero modes.
Actually, the two singlet states are coupled together through $\Delta_q$ and $\Delta_Q$,
whose BdG  equations are derived from Eq. (\ref{HamTot}) such that 
\begin{alignat}1
&
(\hat{\cal H}_0+\hat\Delta_q)(\hat u_r+\hat d_g)+2\hat\Delta_Q \hat s_b=E(\hat u_r+\hat d_g),
\nonumber\\
&
\hat{\cal H}_0\hat s_b+\hat\Delta_Q (\hat u_r+\hat d_g)=E\hat s_b .
\label{BDGEquSin}
\end{alignat}
Here, note that 
any linear combination of singlet states is always singlet, and hence
$\hat u_r+\hat d_g$ and $\hat s_b$ are two singlet states.
Let $((\hat u_r+\hat d_g)/\sqrt{2},\hat s_b)^T$ be an extended wave function. Then, 
we can define a Hamiltonian which leads to Eq.~(\ref{BDGEquSin})
such that
\begin{alignat}1
\left(
\begin{array}{cc}
\hat{\cal H}_0 +\hat\Delta_q& \sqrt{2}\hat\Delta_Q\\
\sqrt{2}\hat\Delta_Q & \hat{\cal H}_0
\end{array}
\right) .
\label{HamSin}
\end{alignat}
This Hamiltonian can be regarded as an effective Hamiltonian in the 
singlet sector.
Chiral decomposition similar to Eq.~(\ref{ChiDecHam}) leads to the Hamiltonian 
\begin{alignat}1
\left(
\begin{array}{cc}
\hat{\cal H}_0^\pm +\hat\Delta_q^\pm& \sqrt{2}\hat\Delta_Q^\pm\\
\sqrt{2}\hat\Delta_Q^\pm & \hat{\cal H}_0^\pm
\end{array}
\right) ,
\end{alignat}
where $\hat{\cal H}_0^\pm$ and $\hat\Delta_Q^\pm$ 
(and $\hat\Delta_q^\pm$ similarly) are defined by Eq.~(\ref{LefRig}).
Let us define new $\Gamma$-matrices 
\begin{alignat}1
&\Gamma^j=\sigma^j\otimes\sigma^3\otimes1, \quad (j=1,2,3)
\nonumber\\
&\Gamma^4=1\otimes\sigma^1\otimes\sigma^1,
\nonumber\\
&\Gamma^5=1\otimes\sigma^1\otimes\sigma^3,
\nonumber\\
&\Gamma^6=1\otimes\sigma^2\otimes1,
\nonumber\\
&\Gamma^7=1\otimes\sigma^1\otimes\sigma^2,
\end{alignat}
which satisfy the relation $\{\Gamma^\mu,\Gamma^\nu\}=2\delta^{\mu\nu}$.
Note that
\begin{alignat}1
&
\Gamma^1\Gamma^2\Gamma^3=i1\otimes\sigma^3\otimes1,
\nonumber\\
&
\Gamma^4\Gamma^5\Gamma^7=-i1\otimes\sigma^1\otimes1,
\nonumber\\
&
\Gamma^5\Gamma^6\Gamma^7=i1\otimes\sigma^2\otimes\sigma^1,
\nonumber\\
&
\Gamma^6\Gamma^4\Gamma^7=i1\otimes\sigma^2\otimes\sigma^3.
\end{alignat}
Then, by the use of the new $\Gamma$ matrices, the above Hamiltonian can be written as
\begin{alignat}1
\hat{\cal H}_{\rm s}^\pm=
\pm\hat{\cal H}_{\rm s}+i\mu\Gamma^1\Gamma^2\Gamma^3 ,
\label{SinChiHam}
\end{alignat}
where
\begin{alignat}1
\hat{\cal H}_{\rm s}&=
\left(
\begin{array}{cccc}
-i\bm\sigma\cdot\nabla &\Delta_q&0&\sqrt{2}\Delta_Q\\
\Delta_q^*& i\bm\sigma\cdot\nabla&\sqrt{2}\Delta_Q^*&0\\
0&\sqrt{2}\Delta_Q&-i\bm\sigma\cdot\nabla&0\\
\sqrt{2}\Delta_Q^*&0&0&i\bm\sigma\cdot\nabla
\end{array}
\right)
\nonumber\\
&=-i\Gamma^j\partial_j+\Gamma^a\phi_a+\frac{i}{2}E^{abc}\psi_a\Gamma^b\Gamma^c\Gamma^7.
\label{SinChiHamCom}
\end{alignat}
Here, two kinds of indices take $j=1,2,3$ and $a=4,5,6$, 
$E^{abc}$ stands for the completely antisymmetric tensor with $E^{456}=+1$, and
\begin{alignat}1
\bm\phi&\equiv(\phi_4,\phi_5,\phi_6)
\nonumber\\
&=\big(\sqrt{2}{\rm Re}\,\Delta_Q,\frac{1}{2}{\rm Re}\,\Delta_q,-\frac{1}{2}{\rm Im}\,\Delta_q\big),
\nonumber\\
\bm\psi&\equiv(\psi_4,\psi_5,\psi_6)
\nonumber\\
&=\big(\sqrt{2}{\rm Im}\,\Delta_Q,\frac{1}{2}{\rm Im}\,\Delta_q,\frac{1}{2}{\rm Re}\,\Delta_q\big).
\label{Hig}
\end{alignat}
The Hamiltonian $\hat{\cal H}_{\rm s}^\pm$ 
thus obtained in (\ref{SinChiHam}) acts on the two singlet states with 
$\gamma_5=\pm1$, respectively.

This Hamiltonian also belongs to the same universality as the previous multiplets, 
class D, since it has particle-hole symmetry
\begin{alignat}1
\widetilde C\hat{\cal H}_{\rm s}^\pm \widetilde C^{-1}=-\hat{\cal H}_{\rm s}^\pm,
\label{ParHolSin}
\end{alignat}
with a reduced particle-hole transformation
\begin{alignat}1
\widetilde C=i\Gamma^1\Gamma^3\Gamma^6K, \quad \widetilde C^2=1.
\end{alignat}

\section{The topological index}\label{s:TopInd} 

In the previous section, we have decomposed the total Hamiltonian 
into several pieces acting on irreducible
multiplets under SU(2)$_{\rm C+F}$ with a definite chirality of
$\gamma_5=\pm1$. These Hamiltonians, denoted as 
$\hat{\cal H}_{m}^\pm$ with $m = $d, t, and s, belong to universality 
class D of topological superconductors. 
We will omit the subscript $m$ for simplicity in this section below.
As discussed by Teo and Kane \cite{TeoKane:10,TeoKane:10b}, 
a line defect in class D allows 
in general protected zero modes of Majorana type 
propagating along a defect.

\subsection{Zero modes along a  vortex}
\label{s:ZeroMode}

To explore such Majorana zero modes,
we assume that the vortex has a cylindrical profile described by 
the cylindrical coordinates $(r,\theta,z)$ such that
\begin{alignat}1
\Delta_q(r,\theta,z)=|\Delta_q(r)|e^{i\Theta_q(\theta)} ,
\label{OrdParPol}
\end{alignat}
where generic angle function $\Theta_q(\theta)$ has a winding $q$, 
$\Theta_q(2\pi)=\Theta_q(0)+2\pi q$. 
We also assume for the radial part that $|\Delta_q(\infty)|=\Delta_{\rm CFL}={\rm const.}>0$ 
as well as $|\Delta_q(0)|=0$ if $q\ne0$ 
($|\Delta_q(0)|'=0$ for $q=0$). 
To investigate the number of gapless modes, we first consider the case of 
the zero chemical potential, $\mu=0$.
For any multiplets, the Hamiltonian is then given by $\pm\hat{\cal H}$,
as can be seen from Eqs. (\ref{DouChiHam}) and (\ref{SinChiHam}).
Since the $\pm$ sign has nothing to do with the zero modes, 
only the Hamiltonian $\hat{\cal H}$ will be henceforth investigated.
The cylindrical symmetry in Eq. (\ref{OrdParPol}) enables us to separate the 
motion to the $z$-direction from others,
\begin{alignat}1
\hat{\cal H}(r,\theta,k_z)=\hat{\cal H}_{\perp}(r,\theta)+k_z\Gamma^3,
\label{HamPer}
\end{alignat}
where the Fourier transformation has been made for the $z$-direction.
Suppose that the eigenvalues for $\hat{\cal H}_{\perp}$ are obtained such that
$\hat{\cal H}_{\perp}\hat\Psi_E=E\hat\Psi_E$. Then, by the use of 
\begin{alignat}1
\{\hat{\cal H}_{\perp},\Gamma^3\}=0 , \quad(\mu=0),
\label{Gam3Chi}
\end{alignat}
we see that 
$\hat{\cal H}_{\perp}(\Gamma^3\hat\Psi_E)=-E(\hat\Gamma^3\hat\Psi_E)$.
We will refer to this property as $\Gamma^3$-chiral symmetry 
of $\hat{\cal H}_{\perp}$.
It thus turns out that the eigenfunctions of $\hat{\cal H}$ 
are linear combinations
of $\hat\Psi_E$ and $\Gamma^3\hat\Psi_E$, and hence, their eigenvalues are those of
\begin{alignat}1
\left(\begin{array}{cc}E&k_z\\k_z&-E\end{array}\right) ,
\nonumber
\end{alignat}
namely, $\pm\sqrt{k_z^2+E^2}$. 
It follows that an $E=0$ state of $\hat{\cal H}_{\perp}$ yields
a zero mode of $\hat{\cal H}$:
Let $\hat\Psi_{0}$ be a zero-energy state of $\hat{\cal H}_\perp$.
Due to $\Gamma^3$-chiral symmetry, we can choose it as 
a simultaneous eigenstate of $\Gamma^3$. 
Suppose $\Gamma^3\hat\Psi_{0\pm}=\pm\hat\Psi_{0\pm}$.
Then, we see that $\hat{\cal H}\hat\Psi_{0\pm}=\pm k_z\hat\Psi_{0\pm}$.
Here, note that the phase of the wave function $\hat\Psi_{0\pm}(r,\theta)$
can be chosen such that
\begin{alignat}1
\widetilde C\hat\Psi_{0\pm}(r,\theta)=\hat\Psi_{0\pm}(r,\theta) ,
\end{alignat}
because of particle-hole symmetry in Eqs. (\ref{ParHolTri}) or (\ref{ParHolSin}).
This tells that the present mode along a vortex line is a chiral Majorana zero mode. 
Therefore, the number of such Majorana modes of $\hat{\cal H}$, 
denoted by $N_{\rm M}(\hat{\cal H})$, is the same 
as the number of zero-energy bound states of $\hat{\cal H}_{\perp}$, 
denoted by $N_0({\cal H}_{\perp})$, which are classified 
by integers, {\it i.e.}, ${\mathbb Z}$.
Namely, 
\begin{alignat}1
N_{\rm M}(\hat{\cal H})=N_0(\hat{\cal H}_{\perp}), \quad (\mu=0).
\end{alignat}

On the other hand, if a nonzero chemical potential $\mu$ is switched on, 
it breaks the $\Gamma^3$-chiral symmetry, since the chemical potential term
in Eqs. (\ref{DouChiHam}) and (\ref{SinChiHam}),
which will be referred to as $\hat{\cal H}_\mu$,
is not anti-commutative with $\Gamma^3$. Let us write the Hamiltonian as
\begin{alignat}1
\hat{\cal H}^\pm(r,\theta,k_z)&=
\underbrace{
\pm\hat{\cal H}_\perp(r,\theta)+\hat{\cal H}_\mu
}
+k_z\Gamma^3
\nonumber\\
&\equiv\hat{\cal H}_{\perp}^\pm(r,\theta)+k_z\Gamma^3,
\label{HamPerChe}
\end{alignat}
and let us consider the zero-energy states of $\hat{\cal H}_\perp^\pm(r,\theta)$.
When $\mu$ is small, we can treat the term $\hat{\cal H}_\mu$ as a perturbation
to the unperturbed degenerate 
zero-energy states of $\hat{\cal H}_\perp$ \cite{FukuiFujiwara:10}.   
We then expect that without any special symmetries,
even number of zero modes of $\hat{\cal H}_\perp$ become 
generically nonzero-energy modes with energies $\pm\varepsilon_j$, 
whereas odd number of zero modes allow at least one unpaired zero mode. 
It may sometimes happen by chance that several zero modes occur if we fine-tune 
parameters of the model, but the evenness or oddness of this number should be invariant.
We can see this alternatively from the symmetry property of the model:
Particle-hole symmetry guarantees that the 
spectrum is symmetric with respect to the zero energy. 
This implies that 
if we consider the spectral flow as a function of $\mu$, the number of zero modes
modulo 2 should be invariant even towards a very large values of $\mu$, 
\begin{alignat}1
N_0(\hat{\cal H}^\pm_\perp)=N_0(\hat{\cal H}_\perp) \mbox{~  mod 2},\quad (\mu\ne0).
\label{NonZerCheNum}
\end{alignat}
Therefore, the above number is a topological invariant characterizing the class D 
superconductors.

Next let us consider the zero modes of $\hat{\cal H}^\pm$. As has been discussed, when $\mu=0$,
we have simultaneous eigenstates of $\hat{\cal H}_\perp^\pm$ and $\Gamma^3$, implying they are
also eigenstates of $\hat{\cal H}^\pm$ for any $k_z$. This is due to $\Gamma^3$-chiral symmetry 
when $\mu=0$.
Contrary to this case, a zero-energy state
of $\hat{\cal H}_\perp^\pm$ is not an eigenstates of $\Gamma^3$ any longer when $\mu\ne0$,
which makes it difficult to find out an eigenstate of $\hat{\cal H}^\pm$
for generic $k_z$.
However, at least when $k_z=0$, it is indeed an eigenstate of $\hat{\cal H}^\pm$
with the zero energy,
and therefore, we can claim the existence of a gapless mode, although we cannot know its exact 
dispersion relation. In Ref. \cite{YasuiIN:10}, an effective theory 
for this gapless mode has been derived. 
Even with a finite chemical potential, the model has particle-hole symmetry
and we can choose wave function with Majorana-like condition.
We thus conclude that the number of gapless Majorana modes of $\hat{\cal H}^\pm$ for $\mu\ne0$
is equivalent to the number of the zero-energy states of $\hat{\cal H}_\perp$ for $\mu=0$
modulo 2,
\begin{alignat}1
N_{\rm M}(\hat{\cal H}^\pm)=N_0(\hat{\cal H}_\perp) \mbox{~  mod 2},\quad (\mu\ne0).
\label{NonZerCheNumMod}
\end{alignat}
In this sense, 
even if we are interested in CFL phase realized at very high density QCD,
it is important to study the case of
the zero chemical potential.
Therefore, we will concentrate on the number of zero-energy states of $\hat{\cal H}_\perp$ in 
each multiplet. 

Now we switch to the discussion on the index of the Hamiltonian.
When $\mu=0$, we can define the following index,
\begin{alignat}1
{\rm ind}\, \hat{\cal H}_\perp=N_+(\hat{\cal H}_\perp)-N_-(\hat{\cal H}_\perp) ,\quad (\mu=0),
\label{AnaInd}
\end{alignat}
where $N_\pm(\hat{\cal H}_\perp)$ is the number of zero-energy states of $\hat{\cal H}_\perp$
with the definite $\Gamma^3$-chirality, $\Gamma^3=\pm1$, respectively.
This is actually possible, since Eq. (\ref{Gam3Chi})  guarantees that 
zero-energy states can be simultaneous eigenstates of $\Gamma^3$.
Since $N_0(\hat{\cal H}_\perp)=N_+(\hat{\cal H}_\perp)+N_-(\hat{\cal H}_\perp)$,
we have $N_0(\hat{\cal H}_\perp)={\rm ind}\,(\hat{\cal H}_\perp)$ mod 2.
Together with (\ref{NonZerCheNum}), we finally reach 
\begin{alignat}1
N_{\rm M}(\hat{\cal H}^\pm)={\rm ind}\,(\hat{\cal H}_\perp)\quad\mbox{mod 2},
\quad (\mu\ne0).
\end{alignat}
Although we have defined the number of zero modes modulo 2,
the typical number is 0 and 1 when the index is even and odd, respectively.
We expect that any other numbers can appear only by chance or due to some symmetries.

\subsection{Index Theorem}
\label{s:Index}

The index theorem states that the index defined above in Eq. (\ref{AnaInd})
can be written by the topological invariant. 
To see this, let us rewrite the index as follows \cite{Weinberg:81,FukuiFujiwara:10};
\begin{alignat}1
{\rm ind}\,\hat{\cal H}_\perp=
\lim_{m\rightarrow0}{\rm Tr}\,\Gamma^3\frac{m^2}{\hat{\cal H}_\perp^2+m^2},
\end{alignat}
where Tr stands for the trace over the 2D coordinate space
as well as over the $\Gamma$-matrices.
In the derivation of the index theorem,  
a central role is played by the axial vector current defined by
\begin{widetext}
\begin{alignat}1
J^j(x,m,M)&=\lim_{y\rightarrow x}
{\rm tr}\,\Gamma^3\Gamma^j
\left(\frac{1}{-i\hat{\cal H}_\perp+m}-\frac{1}{-i\hat{\cal H}_\perp+M}
\right)\delta^2(x-y),
\nonumber\\
&=\lim_{y\rightarrow x}
{\rm tr}\,\Gamma^3\Gamma^j \,(i\hat{\cal H}_\perp)
\left(\frac{1}{\hat{\cal H}_\perp^2+m^2}-\frac{1}{\hat{\cal H}_\perp^2+M^2}
\right)\delta^2(x-y) ,
\label{AxiVecCur}
\end{alignat}
where we have introduced a Pauli-Villars regulator with a large mass parameter $M$ 
to make the current well-defined. After all calculations, 
we should take the limit $M\rightarrow\infty$.
Suppose that $\hat{\cal H}_\perp=-i\Gamma^j\partial_j+\delta\hat{\cal H}_\perp$, 
as in the case above.
Then, the divergence of the current yields the index such that 
\begin{alignat}1
\partial_jJ^j(x,m,M)=2\lim_{y\rightarrow x}{\rm tr}\,\Gamma^3
\left(\frac{m^2}{\hat{\cal H}_\perp^2+m^2}-\frac{M^2}{\hat{\cal H}_\perp^2+M^2}
\right)\delta^2(x-y) .
\end{alignat}
\end{widetext}
Therefore, we reach
\begin{alignat}1
{\rm ind}\,\hat{\cal H}_\perp=c+\lim_{\substack{m\rightarrow0\\M\rightarrow\infty}}\frac{1}{2}
\oint_{|x|\rightarrow\infty} \epsilon_{ij}J^i(x,m,M)dx^j,
\label{IndThe}
\end{alignat}
where
\begin{alignat}1
c=\lim_{M\rightarrow\infty}{\rm Tr}\,\Gamma^3\frac{M^2}{\hat{\cal H}_\perp^2+M^2}.
\end{alignat}
It is known that $c$ becomes the Chern number associated with the gauge potential
\cite{Weinberg:81}.
In the present case, it vanishes, since we have neglected a color magnetic flux confined 
in the vortex.

\subsection{Doublet and triplet states}
\label{s:TopIndTri}

Let us first review the calculations of the topological index, {\it i.e.}, 
the right-hand-side of Eq.~(\ref{IndThe}) 
in the simpler cases of the doublets and triplet, 
which is basically the same as the model already 
studied \cite{Weinberg:81,FukuiFujiwara:10}.

In both cases of doublets and triplet, the effective Hamiltonian $\hat{\cal H}_\perp$
in Eq.~(\ref{IndThe}) is constructed by Eqs.~(\ref{DouChiHamCom}) and (\ref{HamPer})
with Eq.~(\ref{DouOrdPar})
for the doublets and with Eq.~(\ref{TriOrdPar}) for the triplet.
The momentum representation for Eq.~(\ref{AxiVecCur}) yields
\begin{alignat}1
J^j=\int\frac{d^2k}{(2\pi)^2}e^{-ikx}{\rm tr}\,
\Gamma^3\Gamma^j \,(i\hat{\cal H}_\perp)
\frac{1}{\hat{\cal H}_\perp^2+m^2}e^{ikx} ,
\end{alignat}
where $j=1,2$, and contributions from the regulator have been neglected, since 
the above current is well-defined, as seen below.
This is due to the fact that $\Gamma^j$ for $j=1,2$ and the order parameter 
$\Gamma^a\phi_a$ are anti-commutative. 
However, it should be noted that if one considers more generic systems, 
it plays an important role generically.
For convenience, especially for comparison with the singlet sector, we introduce 
the following notations,
\begin{alignat}1
e^{-ikx}(\hat{\cal H}_\perp^2+m^2)e^{ikx}
&=K-\Lambda ,
\end{alignat}
where
\begin{alignat}1
&
K\equiv k^2+\phi^2+m^2,
\nonumber\\
&
\Lambda\equiv i\Gamma^j\Gamma^a\partial_j\phi_a+2ik_j\partial_j+\partial_j^2
\end{alignat}
with $k^2=\sum_{j=1}^2k_j^2$ and $\phi^2=\sum_{a=1}^2\phi_a^2$.
Here $\phi_a$ is given in Eq.~(\ref{DouOrdPar}) for the doublet sectors and
Eq.~(\ref{TriOrdPar}) for the triplet.
Expansion $(K-\Lambda)^{-1}=\sum_n(K^{-1}\Lambda)^nK^{-1}=\sum_n\Lambda^n/K^{n+1}$ leads us to
\begin{alignat}1
J^j&=\sum_{n=0}^\infty
\int\frac{d^2k}{(2\pi)^2}\frac{1}{K^{n+1}}{\rm tr}\,
\Gamma^3\Gamma^j (i\hat{\cal H}_\perp+i\Gamma^\ell k_\ell)
\Lambda^n.
\label{TriCurExp}
\end{alignat}
This expansion is quite useful, since the index theorem (\ref{IndThe}) needs 
the current only
at $|x|\equiv r\rightarrow\infty$: The assumption $|\bm\phi|\rightarrow\Delta_{\rm CFL}
=\mbox{const.}$ 
at $r\rightarrow\infty$ tells that
$\partial_j\phi_a\sim O(r^{-1})$. 
Therefore, we see that the terms $n\ge2$ in Eq.~(\ref{TriCurExp}) vanish when substituted 
into (\ref{IndThe}). 
On the other hand, 
the $n=0$ term also vanishes by the trace. Therefore, only $n=1$ term can contribute to the
index, and we finally reach, by the use of 
${\rm tr}\,\Gamma^3\Gamma^{j}\Gamma^{\ell}\Gamma^a\Gamma^b=(2i)^2\epsilon^{j\ell}\epsilon^{ab}$,
\begin{alignat}1
J^j
&=\int\frac{d^2k}{(2\pi)^2}\frac{1}{K^2}{\rm tr}\,
\Gamma^3\Gamma^ji\Gamma^b\phi_bi\Gamma^\ell\Gamma^b\partial_\ell\phi_b+O(r^{-2})
\nonumber\\
&=(2i)^2\epsilon^{j\ell}\epsilon^{ab}\phi_a\partial_\ell\phi_b\int\frac{d^2k}{(2\pi)^2}\frac{1}{K^2}
\nonumber\\
&=-\frac{1}{\pi(\phi^2+m^2)}\epsilon^{j\ell}\epsilon^{ab}\phi_a\partial_\ell\phi_b .
\end{alignat}
At the infinity $r\rightarrow\infty$, we see 
$\bm\phi/|\bm\phi|=(-\cos\Theta_Q,\sin\Theta_Q)$ for the doublets, and
$\bm\phi/|\bm\phi|=(-\cos\Theta_q,\sin\Theta_q)$ for the triplets.
Taking the limit $m\rightarrow0$,
we finally have
\begin{alignat}1
{\rm ind}\,\hat{\cal H}_\perp=\frac{1}{2\pi}\oint d\theta(\partial_\theta\Theta)=
\left\{
\begin{array}{ll}
Q&(\mbox{doublet})\\q\quad &(\mbox{triplet})
\end{array}
\right.,
\label{TopIndResDou}
\end{alignat}
This result claims that the 2D Hamiltonian $\hat{\cal H}_\perp$ allows at least $Q$ ($q$) zero modes
for each doublet (triplet) state, and therefore, the 3D Hamiltonian with a
vortex line $\hat{\cal H}$ in Eq.~(\ref{HamPer}) has at least $Q$ ($q$) gapless modes along 
the vortex. When the chemical potential is taken into account, it follows from the discussion 
in Sec. \ref{s:ZeroMode} that
the doublet (triplet) allow a zero mode if $Q$ ($q$) is odd.
We have found that 
the non-Abelian $M_1$ vortex ($q=1$ and $Q=0$) has one triplet zero mode 
as explicitly found in \cite{YasuiIN:10},  
while the non-Abelian $M_2$ vortex ($q=0$ and $Q=1$) turns out to have 
one doublet zero mode.

\subsection{Singlet states}

So far we have calculated the index of the 2D Hamiltonian $\hat{\cal H}_\perp$ for both 
the doublet and the triplet states, 
which are basically the same Hamiltonian already studied.
On the other hand, the Hamiltonian $\hat{\cal H}_\perp$ for the singlet states, 
which is defined by Eqs.~(\ref{SinChiHamCom}) and (\ref{HamPer}),
is rather complicated and quite unique, since  
two kinds of different vortices, $\Delta_q$ and $\Delta_Q$, are involved simultaneously. 
In the limiting case where $\Delta_Q=0$, the Hamiltonian 
$\hat{\cal H}_\perp$ is decoupled into 
upper and lower halves. 
The upper is completely the same as the triplet Hamiltonian, whereas the lower is just 
a free fermion Hamiltonian. The index of the former and the latter is $q$ and 0, respectively. 
However, once
the nontrivial $\Delta_Q$ is switched on, 
it is quite interesting to ask which winding,
$\Delta_q$ or $\Delta_Q$,
controls the zero modes of the Hamiltonian.

Although the calculations are parallel to \ref{s:TopIndTri}, the existence of 
the terms composed of three $\Gamma$-matrices in Eq.~(\ref{SinChiHamCom})
makes them quite complicated. 
First, note that
\begin{widetext}
\begin{alignat}1
e^{-ikx}(\hat{\cal H}_\perp^2+m^2)e^{ikx}&=
\left[-i\Gamma^j(\partial_j+ik_j)
+\Gamma^a\phi_a+(i/2)E^{abc}\psi_a\Gamma^a\Gamma^b\Gamma^7\right]^2+m^2
\nonumber\\
&=K+\Gamma-\Lambda,
\end{alignat}
with $j=1,2$ and  $a,b,c=4,5,6$,
where
$\phi_a$ and $\psi_a$ are those in 
Eq.~(\ref{Hig}) for the singlets,
and
\begin{alignat}1
&
K\equiv k^2+m^2+\phi^2+\psi^2,
\nonumber\\
&\Gamma\equiv 2iE^{abc}\Gamma^7\Gamma^a\phi_b\psi_c,
\nonumber\\
&
\Lambda=i\Gamma^j\Gamma^a\partial_j\phi_a
-\frac{1}{2}E^{abc}\Gamma^j\Gamma^b\Gamma^c\Gamma^7\partial_j\psi_a
+2ik_j\partial_j+\partial_j^2 ,
\end{alignat}
with $\phi^2=\sum_{a=1}^3\phi_a^2$ and $\psi^2=\sum_{a=1}^3\psi_a^2$.
Then, the expansion of the kernel $(\hat{\cal H}_\perp^2+m^2)^{-1}$ is also quite useful,
\begin{alignat}1
J^i&=\int\frac{d^2k}{(2\pi)^2}e^{-ikx}{\rm tr}\,\Gamma^3\Gamma^i(i\hat{\cal H}_\perp)
\frac{1}{\hat{\cal H}_\perp^2+m^2}e^{ikx} 
\nonumber\\
&=\int\frac{d^2k}{(2\pi)^2}{\rm tr}\,\Gamma^3\Gamma^i
\left[\Gamma^j(\partial_j+ik_j)+i\Gamma^a\phi_a-(1/2)E^{abc}\psi_a\Gamma^b\Gamma^c\Gamma^7\right]
\frac{1}{K+\Gamma-\Lambda} 
\nonumber\\
&=\sum_{n=0}^\infty
\int\frac{d^2k}{(2\pi)^2}{\rm tr}\,\Gamma^3\Gamma^i
\left[\Gamma^j(\partial_j+ik_j)+i\Gamma^a\phi_a-(1/2)E^{abc}\psi_a\Gamma^b\Gamma^c\Gamma^7\right]
\left(\frac{K-\Gamma}{K^2-\Gamma^2}\Lambda\right)^n\frac{K-\Gamma}{K^2-\Gamma^2} .
\end{alignat}
Here, we have also neglected the regulator with mass $M$.
In the last equation, it is enough to calculate only the $n=1$ term 
for the same reason as in Sec.~\ref{s:TopIndTri}: After taking the trace, we have
\begin{alignat}1
J^j=8\epsilon^{j\ell}
\int\frac{d^2k}{(2\pi)^2}\frac{1}{(K^2-\Gamma^2)^2}
\left\{\left[K^2-2K(\phi^2+\psi^2)+\Gamma^2\right]\phi\cdot\overleftrightarrow{\partial_\ell}\psi
+2K(\phi^2-\psi^2)\overleftrightarrow{\partial_\ell}(\phi\cdot\psi)\right\} ,
\end{alignat} 
where $\phi\cdot\psi=\sum_{a=1}^3\phi_a\psi_a$ and 
$\phi\cdot\overleftrightarrow{\partial_\ell}\psi=\phi\cdot\partial_\ell\psi-(\partial_\ell\phi)\cdot\psi$.
By the use of
$\Gamma^2=4\left(\phi^2\psi^2-(\phi\cdot\psi)^2\right)$ and
the momentum integrations
\begin{alignat}1
&
\int\frac{d^2k}{(2\pi)^2}\frac{K}{(K^2-\Gamma^2)^2}
=\frac{1}{8\pi}\frac{1}{(\phi^2+\psi^2+m^2)^2-4(\psi^2\psi^2-(\phi\cdot\psi)^2)} ,
\nonumber\\
&
\int\frac{d^2k}{(2\pi)^2}\frac{K^2+\Gamma^2}{(K^2-\Gamma^2)^2}
=\frac{1}{4\pi}\frac{\phi^2+\psi^2+m^2}{(\phi^2+\psi^2+m^2)^2-4(\psi^2\psi^2-(\phi\cdot\psi)^2)} ,
\nonumber
\end{alignat}
\end{widetext}
we end up with
\begin{alignat}1
J^j=\epsilon^{j\ell}\frac{2}{\pi}
\frac{
m^2\phi\cdot\overleftrightarrow{\partial_\ell}\psi
+(\phi^2-\psi^2)\overleftrightarrow{\partial_\ell}(\phi\cdot\psi)}
{(\phi^2+\psi^2+m^2)^2-4(\phi^2\psi^2-(\phi\cdot\psi)^2)}.
\label{TriFinCur}
\end{alignat}
When the limit $m\rightarrow0$ is taken, 
the following two cases (1. $\Delta_Q \neq0$ and 2. $\Delta_Q=0$) 
should be considered separately.

\subsubsection{Generic case:  $\Delta_Q\ne0$}

In this generic case, including the case of $Q=0$ (but $\Delta_{Q=0}\ne0$),
we can take the limit $m\rightarrow0$ directly in Eq.~(\ref{TriFinCur}).
Then, we have
\begin{alignat}1
J^j&\rightarrow\epsilon^{j\ell}\frac{2}{\pi}
\frac{
(\phi^2-\psi^2)\overleftrightarrow{\partial_\ell}(\phi\cdot\psi)}
{(\phi^2-\psi^2)^2+4(\phi\cdot\psi)^2}
\nonumber\\
&=
\frac{1}{\pi|\Phi|^2}\epsilon^{j\ell}\epsilon^{\alpha\beta}\Phi_\alpha\partial_\ell\Phi_\beta ,
\end{alignat}
where
\begin{alignat}1
\frac{\bm\Phi}{|\bm\Phi|}
&=\frac{(\phi^2-\psi^2,2\phi\cdot\psi)}{\sqrt{(\phi^2-\psi^2)^2+4(\phi\cdot\psi)^2}}
\nonumber\\
&=(\cos2\Theta_Q,\sin2\Theta_Q).
\end{alignat}
This equation implies that the index of the Hamiltonian is determined solely by the
vortex $\Delta_Q$.
Moreover, the index is twice the winding number of $\Delta_Q$:
\begin{alignat}1
{\rm ind}\,\hat{\cal H}_\perp=\frac{1}{2\pi}\oint d\theta
\partial_\theta(2\Theta_Q) =2Q .
\label{SinTopIndGen}
\end{alignat}
This is the central result in the former part of the present paper:
The topological index of the singlet states is $2Q$ in generic cases.
Therefore, in the typical case of $q=1$ and $Q=0$ (but $\Delta_{Q=0}\ne0$), we conclude that
the index is zero, implying no zero modes generically.
Of course, we cannot deny the existence of even number of zero modes, since the index is
not the number of zero modes. However, in such cases, those zero modes are unstable, and 
small perturbations yield gaps for them \cite{FukuiFujiwara:10}.
In this sense, we expect in general no zero modes in the singlet states.
Actually, in the next Sec. \ref{s:AnaInd}, we solve the differential equation for zero
modes directly, and reach the same conclusion.

We have thus concluded that 
the non-Abelian $M_1$ vortex ($q=1$ and $Q=0$) and $M_2$ vortex ($q=0$ and $Q=1$) 
has no singlet zero modes, but it seems to be inconsistent 
with the result in \cite{YasuiIN:10}, 
in which an asymptotic form of a singlet zero mode 
at large distance from the vortex core was given for the $M_1$ vortex.
In Appendix \ref{s:singlet} we will show that 
the given asymptotic solution 
actually diverges at short distance 
$r\to 0$ around the vortex core. 
We will present another solution which is also well-defined 
at $r\rightarrow\infty$. However, is is diverges as well at short distance.
Therefore those are non-normalizable modes, 
which the index does not count.
We thus conclude that no contradiction exists.

\subsubsection{Exceptional case: $\Delta_Q=0$}

Next, let us consider the exceptional case, $\Delta_Q=0$. 
Although we do not need 
any concrete calculations to know the index as discussed already, 
we show this case below to check the validity of our calculations.
When $\Delta_Q=0$, a naive limit
$m\rightarrow0$ makes the denominator of Eq.~(\ref{TriFinCur}) vanish. However, 
note that $\phi^2=\psi^2$ and $\phi\cdot\psi=0$ when $\Delta_Q=0$. Thus, 
we can take the limit $m\rightarrow0$,
\begin{alignat}1
J^j&=\epsilon^{j\ell}
\frac{2}{\pi}\frac{m^2\phi\cdot\overleftrightarrow{\partial_\ell}\psi}{m^4+4\phi^2m^2}
\nonumber\\
&\rightarrow
-\frac{1}{\pi|\phi|^2}\epsilon^{j\ell}\epsilon^{ab}\phi_a\partial_\ell\phi_b ,
\end{alignat}
where $\bm\phi/|\phi|=(0,\cos\Theta_q,-\sin\Theta_q)$.
Therefore,
\begin{alignat}1
{\rm ind}\,\hat{\cal H}_\perp 
=q .
\label{SinTopIndSpe}
\end{alignat}
We have thus obtained the expected index:
As discussed, when $\Delta_Q=0$, the Hamiltonian decouples into two pieces. 
One is the same as the Hamiltonian for the triplet states, and the other is a free
fermion Hamiltonian. 
It is obvious without any calculations that the former and the latter give 
the index $q$ and 0, respectively.
Once a nonzero $\Delta_Q$ is introduced, 
however small it may be, the index jumps from $q$ to $2Q$,
as seen above.

\section{Analytical index --- Counting zero modes}
\label{s:AnaInd}

So far we have calculated the topological index for the doublet, triplet, and singlet states
as a winding number of the pairing gap function.
In this section, 
we explore normalizable solutions of the equation 
\begin{alignat}1
\hat{\cal H}_\perp(r,\theta)\hat\Psi(r,\theta)=0.
\label{BasZerModEqu}
\end{alignat}
To this end, we assume 
$\Theta_q(\theta)=q\theta$ and $\Theta_Q(\theta)=Q\theta$ in this section.
The merit of such an analysis is that it informs us of
the number of zero-energy states and their $\Gamma^3$-chiralities, $N_+$ and $N_-$, separately.

\subsection{Doublet and triplet states}

For doublet and triplet states, the Hamiltonian $\hat{\cal H}_\perp$ is defined 
by Eqs.~(\ref{DouChiHamCom}) and (\ref{HamPer}). 
In a suitable basis in which $\Gamma^3={\rm diag}(1,1,-1,-1)$, 
Eq.~(\ref{BasZerModEqu}) becomes
\begin{alignat}1
\left(\begin{array}{cc}
0&\hat{\cal H}_{\perp,-}\\
\hat{\cal H}_{\perp,+}&0\end{array}\right)
\left(\begin{array}{cc}\hat\Psi_+\\\hat\Psi_-\end{array}\right)=0,
\label{ChiForH}
\end{alignat}
where $\pm$ denotes the $\Gamma^3$-chirality, $\Gamma^3=\pm1$, and 
\begin{alignat}1
\hat{\cal H}_{\perp,\pm}=
\left(\begin{array}{cc}
-i\partial_\pm&-|\Delta_q(r)|e^{iq\theta}\\
-|\Delta_q(r)|^{-iq\theta}&i\partial_\mp\\
\end{array}\right) ,
\end{alignat} 
with $\partial_\pm=\partial_1\pm i\partial_2=e^{\pm i\theta}(\partial_r\pm\frac{i}{r}\partial_\theta)$.
This is the equation for the triplet, and
if $q$ is replaced by $Q$, it becomes for the doublets. 
We set the $(r,\theta)$-dependence of the wave functions such that
\begin{alignat}1
\hat\Psi_{m\pm}=
\left(\begin{array}{l}
\alpha_m(r)e^{\pm im\theta}\\
i\beta_m(r)e^{ i(\pm m-q\pm 1)\theta}\\
\end{array}\right),
\label{TriDifEqu}
\end{alignat}
where $m$ is a quantum number associated with the angular momentum.
Then, the equation for the radial part is given by
\begin{alignat}1
\left(\frac{d}{dr}-\frac{M_\pm}{r}+\Omega\right)\psi=0,
\label{ZerModEqu}
\end{alignat}
where $\psi=(\alpha_m,\beta_m)^T$, 
\begin{alignat}1
&
M_\pm={\rm diag}\,(m,-m-1\pm q),
\nonumber\\
&
\Omega=
\left(\begin{array}{cc}
0&|\Delta_q|\\
|\Delta_q|&0\\
\end{array}\right) .
\label{MandOme}
\end{alignat}
Counting zero-energy states for this model
has been carried out by Jackiw and Rossi \cite{JacRos81}.
As they have assumed, it may be physically natural to consider the asymptotic gap function
at $r\rightarrow0$ as $|\Delta_q|\sim r^{|q|}$.
We here assume generically
\begin{alignat}1
\Omega=\sum_{n=0}^\infty\Omega_nr^n ,
\label{OmeSer}
\end{alignat}
where $\Omega_0=0$, when $q\ne0$.
From a mathematical point of view, the difference between $|\Delta_q|\sim r^{|q|}$
and $|\Delta_q|\sim r$ does not affect the number of zero-energy states. 
However, the latter is given by log-corrections in the series expansion of the solutions.
See Appendix \ref{s:App} for details. 
Since Eq.~(\ref{ZerModEqu}) is composed of two first-order differential equations,
general solution has two parameters:
One of these parameters is determined by the normalization of the wave function.
As a result, the general solution has {\it one free parameter}.
Let $\psi(r)$ be the general solution of Eq.~(\ref{ZerModEqu}). 
We analyze the asymptotic behavior of $\psi(r)$ below.

First, let us start with the consideration on the 
behavior of $\psi(r)$ at $r\rightarrow\infty$, where we can
neglect the term $1/r$ in Eq.~(\ref{ZerModEqu}). 
The gap function is constant there, 
$|\Delta_q(r)|\rightarrow\Delta_{\rm CFL}~(>0)$. Then, 
the set of equations (\ref{ZerModEqu}) allows two independent solutions 
$\sim e^{\lambda_\pm  r}$,
where $\lambda_\pm=\pm\Delta_{\rm CFL}$ are eigenvalues of $\Omega$.
It follows that the general solution $\psi(r)$ becomes at $r\rightarrow\infty$ a linear combination
of such exponentially decreasing and increasing functions
$\psi(r)\rightarrow\sum_{i=\pm}a_ie^{\lambda_ir}$, where $a_i$ is a constant vector.
For the wave function to be normalizable, we impose {\it one condition} on the wave function 
that $e^{+\Delta_{\rm CFL} r}$ should vanish, $a_+=0$. 
This means that one free parameter mentioned above 
is determined completely. 
One normalizable solution of Eq.~(\ref{ZerModEqu})
thus corresponds to one zero-energy solution for $\hat{\cal H}_\perp$.

Next, let us consider the normalizability of the wave function at $r\rightarrow0$.
For $\Omega$ in Eq.~(\ref{OmeSer}), Eq.~(\ref{ZerModEqu})
can be solved by the series expansion techniques. Details are given in Appendix \ref{s:App}. 
It turns out that two particular solutions are allowed in this limit as well, which are 
written in power series of $r$ such that 
\begin{alignat}1
\psi^{(i)}=\psi_{n_i}^{(i)}r^{n_i}+\psi_{n_i+1}^{(i)}r^{n_i+1}+\cdots ,
\label{SolSerExp}
\end{alignat}
for $i=1,2$, where the leading power $n_i$ is
one of the diagonal elements of $M$, {\it i.e.}, $n_1=m$ and $n_2=-m-1+q$ for $\Gamma^3=+1$ and 
$n_1=m$ and $n_2=-m-1-q$ for $\Gamma^3=-1$.
It should be noted that the general solution $\psi(r)$ becomes a linear combination of $\psi^{(i)}(r)$
in the limit $r\rightarrow0$ and that $\psi$ has no free parameter any longer.
It follows that both $\psi^{(1)}$ and $\psi^{(2)}$ should be normalizable.
The normalizability of $\psi^{(i)}$ in the limit $r\rightarrow0$ is guaranteed by $n_i\ge0$.
Namely, $0\le m\le q-1$ for $\Gamma^3=+1$ and $0\le m\le-q-1$ for $\Gamma^3=-1$.
Accordingly, (1) when $1\le q$, $N_+=q$ and $N_-=0$, (2) when $q=0$, $N_+=N_-=0$, and 
(3) when $q\le -1$, $N_+=0$ and $N_-=-q$. Thus, we end up with 
${\rm ind}\,{\cal H}_\perp=N_+-N_-=q$ in any cases above. 
We have thus reproduced the index in Eq.~(\ref{TopIndResDou}) indeed.

\subsection{Singlet states}

For singlet states, the Hamiltonian $\hat{\cal H}_\perp$ is defined by 
Eqs.~(\ref{SinChiHamCom}) and (\ref{HamPer}). 
In a suitable basis, the zero mode equation becomes 
Eq.~(\ref{ChiForH}) with
\begin{widetext}
\begin{alignat}1
\hat{\cal H}_{\perp,\pm}=
\left(\begin{array}{cccc}
-i\partial_\pm&|\Delta_q(r)|e^{iq\theta}&0&\sqrt{2}|\Delta_Q(r)|e^{iQ\theta}\\
|\Delta_q(r)|^{-iq\theta}&i\partial_\mp&\sqrt{2}|\Delta_Q(r)|e^{-iQ\theta}&0\\
0&\sqrt{2}|\Delta_Q(r)|e^{iQ\theta}&-i\partial_\pm&0\\
\sqrt{2}|\Delta_Q(r)|e^{-iQ\theta}&0&0&i\partial_\mp\\
\end{array}\right) .
\end{alignat} 
\end{widetext}
We set the $(r,\theta)$-dependence of the wave functions such that
\begin{alignat}1
\hat\Psi_{m\pm}=
\left(\begin{array}{l}
\alpha_m(r)e^{\pm im\theta}\\
-i\beta_m(r)e^{ i(\pm m-q\pm 1)\theta}\\
\gamma_m(r)e^{i(\pm m-q+Q)\theta}\\
-i\delta_m(r)e^{i(\pm m-Q\pm 1)\theta}
\end{array}\right).
\end{alignat}
Define $\psi=(\alpha_m,\beta_m,\gamma_m,\delta_m)^T$.
Then, the equation for the radial part is given by the same as Eq.~(\ref{ZerModEqu}),
where 
\begin{alignat}1
&
M_\pm={\rm diag}\,(m,-m-1\pm q,m\pm(Q-q),-m-1\pm Q),
\nonumber\\
&
\Omega=
\left(\begin{array}{cccc}
0&|\Delta_q|&0&\sqrt{2}|\Delta_Q|\\
|\Delta_q|&0&\sqrt{2}|\Delta_Q|&0\\
0&\sqrt{2}|\Delta_Q|&0&0\\
\sqrt{2}|\Delta_Q|&0&0&0\\
\end{array}\right) .
\label{MandOmeSin}
\end{alignat}
We also assume the series expansion Eq.~(\ref{OmeSer}) for $\Omega$.
Counting the zero-energy states will be carried out similarly to the previous case. 
From Eq.~(\ref{MandOmeSin}), it is obvious that when $|\Delta_Q(r)|=0$, Eq.~(\ref{ZerModEqu})
is reduced to two pieces; One corresponds to the triplet states studied above, 
and the other is just free fermions. 
Therefore, in the following 
analysis, we assume $|\Delta_Q(r)|\ne0$ generically.

Equation (\ref{ZerModEqu}) is now composed of four differential equations.
Therefore, in the present case, the general solution includes four parameters,
and one of them is determined by the normalization of the wave function.
The general solution has thus {\it three free parameters}. 
This is the difference between
the present singlet case and previous doublet or triplet cases.
Let $\psi(r)$ be the general solution of Eq.~(\ref{ZerModEqu}).

Let us first consider the asymptotic behavior of $\psi(r)$ 
at $r\rightarrow\infty$, where the $1/r$ term can be 
neglected in Eq.~(\ref{ZerModEqu}). 
We assume that two gap function approach the same constant, 
$|\Delta_q(\infty)|=|\Delta_Q(\infty)|=\Delta_{\rm CFL}={\rm const.}(>0)$.
Then, $\psi(r)$ becomes a linear combination of the 
four independent solutions $\psi\rightarrow \sum_ia_ie^{\lambda_ir}$ with 
$\lambda_{\pm1}=\pm \Delta_{\rm CFL}$ and $\lambda_{\pm2}=\pm 2\Delta_{\rm CFL}$,
where $a_i$ are constant vectors.
For the wave function to be normalizable, we should impose {\it two conditions} 
$a_{+1}=0$ and $a_{+2}=0$.
Therefore, the general solution has still {\it one free parameter}.
This is in sharp contrast to the previous case in which the general solution has 
no free parameters at this stage.

Let us next consider the behavior of $\psi(r)$ at $r\rightarrow0$. 
In this region, we can solve 
Eq.~(\ref{ZerModEqu}) by using the power series expansion of the wave functions. 
For details, see Appendix \ref{s:App}.
We show there that the set of four equations allow four special solutions
(\ref{SolSerExp}) 
whose leading power $n_i$ $(i=1,\cdots,4)$
is given by the diagonal elements of $M_\pm$ defined by Eq.~(\ref{MandOmeSin}).
It follows that, near $r\rightarrow0$, we have
\begin{alignat}1
\psi\rightarrow\sum_{i=1}^4c_i\psi^{(i)} ,
\label{LinComSol}
\end{alignat}
where $c_i$ is a constant.
When we consider the normalizability of the wave function $\psi(r)$, 
one free parameter plays a crucial role, since {\it one of $n_i$ can be negative}.
In what follows, we restrict our discussions to $\Gamma^3=+1$ case, for simplicity.

Suppose $Q\ge0$. 
Then, the four powers are given by $n_1=m$, $n_2=-m-1+q$, $n_3=m+Q-q$ and
$n_4=-m-1+Q$ for $\Gamma^3=+1$ states. We should have some cases separately.

\begin{itemize}
\item[(I)] Case of $Q\le q$.
\begin{itemize}
\item[(1)] If all $n_i$ are non-negative, $0\le n_{1,2,3,4}$, 
we have $q-Q\le m\le Q-1$. In this case,
all $\psi^{(i)}$ $(i=1,\cdots,4)$ are normalizable at $r\rightarrow0$.
This implies that we do not need to impose any conditions of normalizability on $\psi(r)$.
Remember here that the general solution $\psi(r)$ has {\it one free parameter}.
Therefore, $\psi(r)$ is a linear combination of two independent solutions.
In other words, this case allows degenerate two solutions for each quantum number $m$.
Therefore, we have
\begin{alignat}1
N_+=\left\{\begin{array}{ll}2(2Q-q)\quad&(q\le2Q)\\0&(2Q\le q)\end{array}\right. ,
\end{alignat}
where, the factor 2 above means two independent solutions.
\item[(2)] If $n_1<0$, $0\le n_{2,3,4}$, no $m$ is possible.
\item[(3)] If $n_2<0$, $0\le n_{1,3,4}$, no $m$ is possible.
\item[(4)] If $n_3<0$, $0\le n_{1,2,4}$, we have
\begin{alignat}1
0\le m\le \left\{\begin{array}{ll}q-Q-1\quad&(q\le 2Q)\\Q-1 &(2Q\le q)\end{array}\right. .
\end{alignat}
Since $n_3<0$, this particular solution should be eliminated from $\psi(r)$
at $r\rightarrow0$ given in Eq.~(\ref{LinComSol}) 
by imposing the additional condition $c_3=0$. 
This determines one free parameter which $\psi(r)$ has, and normalizable $\psi(r)$ thus obtained
gives one solution for each quantum number $m$. Therefore,
\begin{alignat}1
N_+=\left\{\begin{array}{ll}q-Q\quad&(q\le2Q)\\ Q&(2Q\le q)\end{array}\right..
\end{alignat}
\item[(5)] If $n_4<0$, $0\le n_{1,2,3}$, we have
\begin{alignat}1
\left\{\begin{array}{l}Q\\q-Q\end{array}\right.\le m\le q-1 \quad
\begin{array}{l}(q\le2Q)\\(2Q\le q)\end{array} .
\end{alignat}
For the same reason as in the case of (4),
for each allowed $m$, one zero-energy state is given, 
and the number of the zero-energy states are thus
\begin{alignat}1
N_+=\left\{\begin{array}{ll}q-Q\quad&(q\le2Q)\\ Q&(2Q\le q)\end{array}\right..
\end{alignat}
\end{itemize}
Summing up $N_+$ over the cases from (1) to (5), we conclude that  
the number of zero-energy state is given by $N_+=2Q$
in the case of $Q\le q$.
\item[(II)] Case of $q\le Q$.
\begin{itemize}
\item[(1)] If $0\le n_{1,2,3,4}$, 
we have $0\le m\le q-1$. Since no additional conditions of the normalizability 
need to be imposed, 
there are two degenerate solutions for each $m$.
Therefore, we have
\begin{alignat}1
N_+=\left\{\begin{array}{ll}2q\quad&(0\le q)\\0&(q\le 0)\end{array}\right. .
\end{alignat}
Note here the factor 2, as discussed in (I) (1).
\item[(2)] If $n_1<0$, $0\le n_{2,3,4}$, we have
\begin{alignat}1
q-Q\le m\le \left\{\begin{array}{ll}-1\quad&(0\le q)\\q-1 &(q\le 0)\end{array}\right. .
\end{alignat}
Since each allowed $m$ gives one zero-energy state, 
and the number of the zero-energy state is given by
\begin{alignat}1
N_+=\left\{\begin{array}{ll}Q-q\quad&(0\le q)\\ Q&(q\le 0)\end{array}\right. .
\end{alignat}
\item[(3)] If $n_2<0$, $0\le n_{1,3,4}$, we have
\begin{alignat}1
\left\{\begin{array}{l}q\\0\end{array}\right.\le m\le Q-1 \quad
\begin{array}{l}(0\le q)\\(q\le 0)\end{array}.
\end{alignat}
The number of the zero-energy state is
\begin{alignat}1
N_+=\left\{\begin{array}{ll}Q-q\quad&(0\le q)\\ Q&(q\le 0)\end{array}\right. .
\end{alignat}
\item[(4)] If $n_3<0$, $0\le n_{1,2,4}$, no $m$ is possible.
\item[(5)] If $n_4<0$, $0\le n_{1,2,3}$, no $m$ is possible.
\end{itemize}
It turns out that the number of zero-energy states is 
also given by $N_+=2Q$ in this case.
\end{itemize}
So far we have counted the zero-energy states with $\Gamma^3=+1$ in the case of $Q\ge0$.
In a similar way, we immediately see that there are no solutions 
allowed in the $\Gamma^3=-1$ states.
Thus we conclude that $N_+=2Q$ and $N_-=0$ when $Q\ge0$.
On the other hand, in the case of $Q\le0$, we can show that $N_+=0$ and $N_-=2|Q|$.
These results match precisely the topological index given by Eq.~(\ref{SinTopIndGen}).

In the above analysis, we have used just the leading power of $\psi_i(r)$.
In Appendix \ref{s:App}, we discuss the higher orders of the series expansion
and show how to construct the solutions. 
This formulation may be interesting on its own right, 
in particular, in that we need to introduce
$(\ln r)^n$ $(n\le3)$ terms to obtain four independent solutions.

\section{Summary and discussions}\label{s:Sum} 
Non-Abelian vortices exist in the CFL phase
of color superconductivity.
We have calculated the topological and analytical indices 
for fermion zero modes of non-Abelian vortices 
$\Delta = {\rm diag}(\Delta_Q,\Delta_Q,\Delta_q)$ 
which preserve SU(2)$_{\rm C+F}$ symmetry in their core,
where $\Delta_Q$ and $\Delta_q$ have winding $Q$ and $q$, respectively.
Because of SU(2)$_{\rm C+F}$ symmetry unbroken in the core of 
the vortex, the BdG Hamiltonian is decomposed into some sectors 
according to the irreducible representations 
of SU(2)$_{\rm C+F}$; triplet, doublet and singlet sectors.
In the limit of the zero chemical potential $\mu=0$, 
the topological indices for the triplet, doublet and singlet zero modes 
of SU(2)$_{\rm C+F}$ 
have been calculated to be $q$, $Q$ and $2Q$, respectively.
We have also analyzed the BdG equations, and from the normalizability of the 
wave function we have obtained the same indices of the zero modes.
Because of particle-hole symmetry, 
these zero modes can be regarded as chiral Majorana fermion modes.
For non-zero chemical potential $\mu \neq 0$, 
we have expected that the triplet or doublet zero modes exist 
only when $q$ or $Q$ is odd, respectively, whereas no zero modes in the singlet states.
Technically speaking, 
the derivation of the indices of the triplet and doublet states 
is essentially the same with that of one component \cite{Weinberg:81,FukuiFujiwara:10}, 
but that of the singlet states is quite nontrivial;
the index of the singlets is $q$ if we set $\Delta_Q$ to zero by hand,  
but it jumps to $2Q$ once non-zero $\Delta_Q$ is turned on,
however small it may be.

In particular,  as the most fundamental non-Abelian vortices, 
the non-Abelian $M_1$ vortex ($q=1$ and $Q=0$)
carrying 1/3 quantized circulation of $U(1)_{\rm B}$,
has 
one triplet, no doublet and no singlet Majorana zero modes. 
Although the number of triplets agrees with 
the previous result in \cite{YasuiIN:10}, 
that of singlet does not seem so at first glance. 
However we have shown in Appendix \ref{s:singlet} 
that the asymptotic zero mode at large distance from the vortex core
found in \cite{YasuiIN:10} diverges in the core.
The index does not count such a non-normalizable mode,
and hence, no inconsistency exists.
On the other hand, 
the non-Abelian $M_2$ vortex ($q=0$ and $Q=1$) 
carrying 2/3 quantized circulation of $U(1)_{\rm B}$, 
has no triplet, one doublet and no singlet Majorana zero modes.

\bigskip
Before closing this paper, 
several discussions are addressed here.

We have calculated the indices for non-Abelian vortices
of a composite of $q$ $M_1$ and $Q$ $M_2$ types, where 
the SU(2)$_{\rm C+F}\times$U(1)$_{\rm C+F}$ symmetry remains 
in the core of vortices.
Generalization to the case with three components 
having all different winding numbers is straightforward, 
in which case only U(1)$_{\rm C+F}^2$ symmetry remains.

Although we have considered (composite) vortices 
placed in the same position, 
they can be separated if they are composite.
In fact, a non-Abelian $M_2$ vortex can be decomposed 
into two non-Abelian $M_1$ vortices winding in different components.
It is an interesting problem to see 
how the indices of SU(2) multiplets change 
from $M_2$ to $M_1$'s in this process.
Here,  the positions of the vortices play a crucial role 
as a new degree of freedom.
One doublet zero mode of one $M_2$ vortex will disappear, as it separates 
into two $M_1$ vortices, and when they are well separated in position, 
a triplet zero mode will appear in each of two $M_1$ vortices. They are at 
the zero energy 
when two $M_1$ vortices are infinitely separated. 
Thus, the doublet zero modes are lifted to non-zero modes in 
a process of decomposition of one $M_2$ vortex into two $M_1$ vortices,
whereas new zero modes appear in turn in the triplet states.
It implies the existence of attractive and repulsive 
forces between two $M_1$ vortices, 
mediated by doublet and triplet (and maybe singlet) fermion modes, respectively.

A new non-Abelian statistics of 
non-Abelian $M_1$ vortices with triplet zero modes 
was studied \cite{Yasui:2010yh},
in which it was shown to be written as a tensor product of 
that of Abelian vortices in a chiral $p$-wave superconductor \cite{Ivanov:01} 
and the Coxter group.
Our result implies that 
non-Abelian $M_2$ vortices have doublet zero modes, 
which may give another new non-Abelian statistics.

Non-Abelian vortices should exist in the core of neutron stars 
if the CFL phase is indeed realized there.
They will constitute a vortex lattice 
because of the rapid rotation of the stars
\cite{Nakano:2007dr,Sedrakian:2008ay}.
It is known that fermion zero modes on vortex lines 
significantly change the transport properties of quasi-particles \cite{Volovik}.
Fermion zero modes found in this paper will be essential 
in the study of dynamics of neutron stars, 
which may hopefully gives an observational signal.

Finally, the CFL phase admits 
a variety of topological defects; 
domain walls, Skyrmions \cite{Hong:1999dk},
 (confined) monopoles \cite{Gorsky:2011hd} 
and instantons \cite{Schafer:2002ty}.  
A generalization to the index theorems 
in the presence of these topological defects  
will be an interesting work to be explored.
It also counts edge modes at the phase boundary \cite{Sadzikowski:2002in,Giannakis:2003am} 
between the CFL phase and the hadron phase, if it exists. 
Also, the inclusion of strange quark masses \cite{Eto:2009tr} 
and an extension to the other phases such as 2SC phase \cite{Alford:2007xm} 
will be important in application to more realistic situation.

\begin{acknowledgments}
This work was supported in part by Grants-in-Aid for Scientific Research from JSPS
(Nos. 20340098 and 21540378) for Fukui and 
Grants-in-Aid for Scientific Research 
(Nos. 20740141 and 23740198) from the Ministry of Education, Culture, Sports, Science 
and Technology-Japan for Nitta.
This work was also supported by the ``Topological Quantum Phenomena'' 
(No. 23103502 for Fukui No. 23103515 for Nitta) Grant-in Aid for Scientific Research 
on Innovative Areas from the Ministry of Education, Culture, Sports, Science and Technology
(MEXT) of Japan.
Yasui is supported by the Grant-in-Aid for Scientific Research on Priority
Areas ``Elucidation of New Hadrons with a Variety of
Flavors'' (No. 21105006) from the MEXT of Japan.

\end{acknowledgments}

\appendix

\section{Singlet solutions in the asymptotic form at large $r$}\label{s:singlet}
\label{asympt:App}

The asymptotic form of
a singlet zero mode of a non-Abelian $M_1$ vortex at large $r$  
was given in \cite{YasuiIN:10}.
This is actually a candidate of zero-energy states, which is exponentially decreasing function 
at $r\rightarrow\infty$. 
However, the asymptotics at $r\rightarrow0$ is also crucial for
the normalizability of wave functions. 
To investigate the behavior of this solution extrapolated into a small $r$ region,
we need to solve the BdG equation numerically.

For numerical calculations with high accuracy, it is desirable that
we have two independent solutions at $r\rightarrow\infty$. 
A linear combination of them is then a general solution including a parameter.
%
Starting with such a generic wave function, we can investigate precisely the behavior of the 
numerical wave function extrapolated into
$r\rightarrow0$ by changing the parameter of the generic wave function 
at the boundary $r\rightarrow\infty$. 

To this end, we will present not only the one shown in Ref. \cite{YasuiIN:10}
but also another asymptotic solution at $r\rightarrow\infty$ in this Appendix, 
and report the result of the numerical computation of the BdG equation. 


We show the explicit form of the wave functions of the singlet solution for the right mode ($\gamma_{5}=+1$) in the Weyl representation
\begin{eqnarray}
\hat{u}_{r} \!=\!
\left(
\begin{array}{c}
 \varphi_{1}(r,\theta) \\
 \eta_{1}(r,\theta)
\end{array}
\right), \hspace{1em}
\hat{d}_{g} \!=\!
\left(
\begin{array}{c}
 \varphi_{2}(r,\theta) \\
 \eta_{2}(r,\theta)
\end{array}
\right),
\end{eqnarray}
and
\begin{eqnarray}
\hat{s}_{b} \!=\!
\left(
\begin{array}{c}
 \varphi_{3}(r,\theta) \\
 \eta_{3}(r,\theta)
\end{array}
\right),
\end{eqnarray}
where
\begin{eqnarray}
\varphi_{i}(r,\theta) =
\left(
\begin{array}{c}
 f_{i}(r) \\
 i g_{i}(r) e^{i\theta} \\
 0 \\
 0
\end{array}
\right), \\
\eta_{i}(r,\theta) =
\left(
\begin{array}{c}
 0 \\
 0 \\
 \bar{f}_{i}(r) e^{-i\theta} \\
 i \bar{g}_{i}(r)
\end{array}
\right),
\end{eqnarray}
for $\hat{u}_{r}$ ($i=1$) and $\hat{d}_{g}$ ($i=2$), and
\begin{eqnarray}
\varphi_{3}(r,\theta) &=&
\left(
\begin{array}{c}
 f_{3}(r) e^{-i\theta} \\
 i g_{3}(r) \\
 0 \\
 0
\end{array}
\right), \\
\eta_{3}(r,\theta) &=&
\left(
\begin{array}{c}
 0 \\
 0 \\
 \bar{f}_{3}(r) \\
 i \bar{g}_{3}(r) e^{i\theta}
\end{array}
\right),
\end{eqnarray}
for $\hat{s}_{b}$. 
Here we have the relations 
$\bar{f}_{1}(r) = - g_{2}(r)$, $\bar{g}_{1}(r) = f_{2}(r)$, $\bar{f}_{2}(r) = - g_{1}(r)$, 
$\bar{g}_{1}(r)= f_{2}(r)$, $\bar{f}_{3}(r) = - g_{2}(r)$ and $\bar{g}_{3}(r)= f_{3}(r)$ 
from the Majorana condition.
At large $r$, both $|\Delta_{0}|$ and $|\Delta_{1}|$ become a common constant $|\Delta|$ 
given in the bulk state, $|\Delta_0|,~|\Delta_1|\rightarrow|\Delta|\equiv \Delta_{\rm CFL}(>0)$ 
at $r\rightarrow\infty$.
Then, with an approximation of small $|\Delta|$ and large $\mu$, we find asymptotic forms 
of the wave functions with a condition of the convergence at large $r$.
The first solution is
\begin{eqnarray}
f_{i}(r) &=& {\cal N} e^{-|\Delta|r/2} J_{0}(\mu r), \\
g_{i}(r) &=& {\cal N} e^{-|\Delta|r/2} J_{1}(\mu r),
\end{eqnarray}
($i=1$, $2$) and
\begin{eqnarray}
f_{3}(r) &=& - \frac{{\cal N}}{2} e^{-|\Delta|r/2} J_{0}(\mu r), \\
g_{3}(r) &=& - \frac{{\cal N}}{2} e^{-|\Delta|r/2} J_{1}(\mu r),
\end{eqnarray}
with a normalization constant ${\cal N}$.
This is the solution given in the previous work \cite{YasuiIN:10}.
As a second solution, we find a new asymptotic solution, 
which was not considered in the previous work,
\begin{eqnarray}
\hspace{-0.5em} f'_{i}(r) &\!=\!& {\cal N}' e^{-|\Delta|r/2} \frac{\pi}{4} (\mu r)^{2} J_{1}(\mu r) \nonumber \\
 &&\hspace{1em} \big( J_{1}(\mu r) N_{0}(\mu r) - J_{0}(\mu r) N_{1}(\mu r) \big), \\
\hspace{-0.5em} g'_{i}(r) &\!=\!& {\cal N}' e^{-|\Delta|r/2} \frac{1}{4} \big( -\mu r J_{0}(\mu r) + J_{1}(\mu r) \big),
\end{eqnarray}
($i=1$, $2$) and
\begin{eqnarray}
\hspace{-0.5em} f'_{3}(r) &\!=\!& {\cal N}' e^{-|\Delta|r/2} \frac{\mu}{8|\Delta|} \Big\{ 4 \left( \mu r J_{0}(\mu r) - J_{1}(\mu r) \right) \nonumber \\
&&  - \pi (\mu r)^{2} \left( 2 J_{0}(\mu r) + \frac{|\Delta|}{\mu} J_{1}(\mu r) \right) \nonumber \\
&& \hspace{0.8em} \big( J_{1}(\mu r) N_{0}(\mu r) - J_{0}(\mu r) N_{1}(\mu r) \big) \Big\}, \\
\hspace{-0.5em} g'_{3}(r) &\!=\!& {\cal N}' e^{-|\Delta|r/2} \frac{\mu}{4|\Delta|} \Big\{ \left( 2+|\Delta|r \right) J_{0}(\mu r) \nonumber \\
&& - \left( \frac{|\Delta|}{\mu} + 2\mu r \right) J_{1}(\mu r) \nonumber \\
&&  + \pi (\mu r)^{2} J_{1}(\mu r) \nonumber \\
&& \hspace{0.8em} \big( J_{1}(\mu r) N_{0}(\mu r) - J_{0}(\mu r) N_{1}(\mu r) \big) \Big\},
\end{eqnarray}
with a normalization constant ${\cal N'}$.
It should be emphasized that these asymptotic solutions are correct 
only at large $r$, at which $|\Delta_{0}|$ and $|\Delta_{1}|$ are constant.
However, these solutions may be divergent in small $r$ in general, 
because $|\Delta_{1}|$ becomes zero at $r=0$, 
and we find that it is the case.
In order to see the behavior at small $r$, we have solved numerically the BdG equation 
with assuming an approximate $r$-dependence of $\Delta_{0}(r)$ and $\Delta_{1}(r)$ 
(for example, $\Delta_{0}(r)=\Delta_{\rm CFL}= \mbox{const.}>0 $
and $\Delta_{1}(r)=\Delta_{\rm CFL} \tanh \xi r$ 
with a coherence length $\xi$) with starting from large $r$ in which these asymptotic solutions 
are given as a boundary condition.
As a numerical result, we have found that the wave functions become divergent at $r=0$ 
for any linear combinations of the first and the second asymptotic solutions as a boundary condition.
It means that these asymptotic solutions are not normalizable in the whole range of $r$.
Therefore it is confirmed that there is no normalizable zero-energy state in the singlet, 
consistent with the result of the index theorem in the text.

\section{Analytic solution of the differential equations}
\label{s:App}

We have counted the zero modes based on a simple assumption that
the leading power of the particular solutions for Eq.~(\ref{ZerModEqu}) near $r\sim0$
is given by the diagonal elements of the matrix $M_\pm$ in Eqs. (\ref{MandOme}) and
(\ref{MandOmeSin}). In this Appendix we show it indeed possible 
to find such solutions.

We assume that the differential equation 
has a power series solution 
\begin{alignat}1
  \psi=r^\lambda\sum_{n=0}^\infty\psi_{\lambda+n} r^n. 
\end{alignat}
Putting this as well as the Taylor expansion of $\Omega$ in Eq.~(\ref{OmeSer}) into 
the differential equation Eq.~(\ref{ZerModEqu}), 
we can write the differential equation as
\begin{widetext}
\begin{alignat}1
  (\lambda-M)\psi_{\lambda}r^{\lambda-1}
  +\sum_{n=0}^\infty\left\{(\lambda+n+1-M)\psi_{\lambda+n+1}
  +\sum_{l=0}^n\Omega_{n-l}\psi_{\lambda+l}\right\}r^{n+\lambda}=0,
\end{alignat}
\end{widetext}
where $M_\pm$ has been denoted simply as $M$.
This gives
\begin{alignat}1
  &(\lambda-M)\psi_\lambda=0, 
\label{LeaDifEqu}\\ 
  &(\lambda+n+1-M)\psi_{\lambda+n+1}
  +\sum_{l=0}^n\Omega_{n-l}\psi_{\lambda+l}=0,
\label{NexDifEqu}
\end{alignat}
with $n\ge0$ for the latter. From Eq.~(\ref{LeaDifEqu}), 
it follows that $\lambda$ should be one of the diagonal
elements of $M$.
For the largest value of $\lambda$, Eq.~(\ref{NexDifEqu}) can be solved recursively,
whereas for smaller $\lambda$, we could  meet the difficulty, since at a certain
$n$, $\det (\lambda+n+1-M)=0$.

\subsection{Case of triplet and doublets}

Without loss of generality, we can set
\begin{alignat}1
M={\rm diag}(p,s), \quad p\ge s,
\end{alignat}
where $p$ and $s$ are integers.
As mentioned above, the leading power $\lambda$ is given by 
$\lambda=p$ and $\lambda=s$. 
Correspondingly, Eq.~(\ref{LeaDifEqu}) gives
two solutions $\psi^{(i)}_\lambda$ ($i=1,2$) as 
\begin{alignat}1
  \psi^{(1)}_p=\left(
    \begin{matrix}
      1 \\ 0
    \end{matrix}\right), \quad
  \psi^{(2)}_{s}=
  \left(\begin{matrix}
    0 \\ 1
  \end{matrix}\right). 
\label{IniStaDou}
\end{alignat}
If $p=s$, we see $\det(n+\lambda+1-M)\ne0$ for $n\geq0$. 
We can find 
$\psi^{(i)}_{n+\lambda+1}$ recursively 
by (\ref{NexDifEqu}). We thus obtain two solutions.  
Henceforth,  we assume $p>q$. 

\subsubsection{Generic case of $|\Delta_q|\sim r$}

First of all, let us consider the most generic case and give the solutions 
of Eq.~(\ref{ZerModEqu}) just by assuming Eq.~(\ref{OmeSer}).
For $\lambda=p$, 
higher $\psi^{(1)}_{p+n+1}$ $(n\ge0)$ can be obtained 
recursively by Eq.~(\ref{NexDifEqu}), and the solution is given by
\begin{alignat}1
  \psi^{(1)}(r)=\sum_{n=0}^\infty
  \psi_{n+p}^{(1)}r^{n+p}.
  \label{FiSolDC}
\end{alignat}

For the smaller power $\lambda=s$, the recursion relation is ill-defined at 
$n=p-s-1\ge0$, since $\det(p-M)=0$. This has a close relationship with 
the well-known fact that the Bessel function $J_n$ and $J_{-n}$ are not
independent when $n$ is an integer, and the Neumann function $Y_n$ in addition to the Bessel 
function $J_n$ can be regarded as two independent solutions of Bessel's differential equation.
In the present problem, 
the standard technique of solving differential equations leads to the following ansatz 
\begin{alignat}1
  \psi^{(2)}(r)=\sum_{n=0}^\infty
  \psi_{n+s}^{(2)}r^{n+s}+a\psi^{(1)}(r)\ln r, 
\label{TriSecSol}
\end{alignat}
where $a$ is an unknown constant to be determined. Substituting this into 
Eq.~(\ref{ZerModEqu}) modifies Eq.~(\ref{NexDifEqu}) such that 
\begin{alignat}1
  &(s+n+1-M)\psi^{(2)}_{s+n+1}
  +\sum_{l=0}^n\Omega_{n-l}\psi^{(2)}_{s+l}+a\psi^{(1)}_{s+n+1}=0,
\end{alignat}
where $\psi^{(1)}_{n+s+1}=0$ for $n=0,\cdots,p-s-2$ is assumed. 
The difficulty at $n=p-s-1$ due to $\det(p-M)=0$ is 
avoided by introducing a parameter $a$. 
We can determine both $a$ and $\psi^{(2)}_{p}$ by this equation. 
For details, especially for the uniqueness of the solution,
see the discussions in App. \ref{s:Two}.
Once these are obtained, we solve $\psi_{n+s+1}^{(2)}$ for 
$n\geq p-s$ recursively. We thus obtain the 
second solution (\ref{TriSecSol}). Although it contains $\ln r$, 
the expected leading power behaviors near the origin
is not altered.

\subsubsection{Case of $|\Delta_q|\sim r^{|q|}$}

So far we have shown that if the diagonal elements of $M$ are different integers,
the solutions include generically a $\ln r$ term.
We will show below 
that if we assume $|\Delta_q|\sim r^{|q|}$, the normalizable solutions of 
Eq.~(\ref{ZerModEqu}) do not include such a term, and they can be 
given by purely power series of $r$.

We restrict our discussions to $\Gamma^3=+1$ and $q\ge1$, for definiteness.
Then,
\begin{alignat}1
M=M_+={\rm diag}(m,q-m-1) ,
\end{alignat} 
and normalizability requires $0\le m\le q-1$. 
We explore the case $m>q-m-1$ only. The solution for the larger $\lambda=m$ 
is given by (\ref{FiSolDC}).
For the smaller $\lambda=q-m-1$, we can choose $\psi_{q-m-1}^{(2)}$ to be 
the second one in Eq.~(\ref{IniStaDou}).
Eq.~(\ref{NexDifEqu}) gives recursively 
$\psi_{n+q-m}^{(2)}$ for $0\leq n<2m-q$. 
The problematic equation at $n=2m-q$ 
can be satisfied simply by choosing 
$\psi_{m}^{(2)}=0$, since couplings with 
$\Omega_q$ only appear in (\ref{NexDifEqu}) 
for $n\geq q$, whereas $0\le 2m-q\le q-2$.
Eq.~(\ref{NexDifEqu}) then 
determines $\psi_{n+q-m}$ for $n> 2m-q$, 
giving the second solution
\begin{alignat}1
  \psi^{(2)}(r)=\sum_{n=0}^\infty
  \psi_{n+q-m-1}^{(2)}r^{n+q-m-1}.
\end{alignat}
Therefore, in this case, there are no terms including $\ln r$.

\subsection{Case of singlet}

This case is rather complicated, since Eq. (\ref{ZerModEqu}) is composed of 
four coupled equations. 
We first rearrange the differential equation 
Eq.~(\ref{ZerModEqu}) for the singlet case to satisfy 
\begin{alignat}1
M={\rm diag}(n_1,n_2,n_3,n_4),
\nonumber\\
\quad n_1\geq n_2 \geq n_3\geq n_4 ,
\end{alignat}
and introduce a set of normalized eigenvectors 
$\psi^{(i)}_{n_i}$ ($i=1,\cdots,4$) by 
\begin{alignat}1
  \psi_{n_1}^{(1)}=\left(
    \begin{matrix}
      1\\0\\0\\0
    \end{matrix}\right), \quad\cdots, \quad
  \psi_{n_4}^{(4)}=\left(
    \begin{matrix}
      0\\0\\0\\1
    \end{matrix}\right),
  \label{NBases}
\end{alignat}
The solutions of Eq.~(\ref{ZerModEqu}) near the 
origin change their forms if any of $n_i$ are 
equal. Such degeneracies are classified into 
eight different cases. 
We consider these separately.  
In particular, when there are more than two different diagonal elements, 
there appear $(\ln r)^n$ corrections with $n\ge2$, which is a new feature absent 
in the previous triplet and doublet cases.

\subsubsection{$n_1=n_2=n_3=n_4$}
\label{s:One}

We define $p=n_1$. 
Eq.~(\ref{LeaDifEqu}) gives four independent solutions 
$\psi_p=\psi_p^{(i)}$ ($i=1,\cdots,4$) for $\lambda=p$. 
Eq.~(\ref{NexDifEqu}) then recursively determine 
$\psi_{p+n+1}=\psi_{p+n+1}^{(i)}$. We thus obtain 
four solutions given by
\begin{alignat}2
  \psi^{(i)}(r)=\sum_{n=0}^\infty\psi^{(i)}_{n+p}r^{n+p} ,
  \quad(i=1,\cdots,4) .
\end{alignat}

\subsubsection{$n_1=n_2=n_3>n_4$ }
\label{s:Two}

Let us define $p=n_1$ and $s=n_4$. Then, 
three solutions $\psi^{(i)}$ ($i=1,2,3$) with 
$\lambda=p$ can be obtained as in the case of App. \ref{s:One}.
The fourth solution for $\lambda=s$ can be assumed to be
\begin{alignat}1
  \psi^{(4)}=\sum_{n=0}^\infty\psi^{(4)}_{n+s}r^{n+s}
  +\sum_{i=1,2,3}a^{(4,i)}\psi^{(i)}(r)\ln r, 
  \label{CiiFoSol}
\end{alignat}
where 
$a^{(4,i)}$ are some constants. The recursion 
relations for the coefficients can be found by 
inserting Eq.~(\ref{CiiFoSol}) into Eq.~(\ref{ZerModEqu}). 
They read
\begin{alignat}1
  &(s-M)\psi^{(4)}_s=0, 
  \label{CiiIc} \\
  &(s+n+1-M)\psi_{s+n+1}^{(4)}+\sum_{l=0}^n
  \Omega_{n-l}\psi^{(4)}_{s+l}  \nonumber \\
  &\hskip .5cm +\sum_{i=1,2,3}a^{(4,i)}\psi^{(i)}_{s+n+1}=0, 
  \quad(n=0,1,\cdots)
  \label{CiiRecEq}
\end{alignat}
where $\psi^{(1,2,3)}_{s+n}=0$ for $n=0,1,\cdots,p-s-2$ is 
assumed. Eq.~(\ref{CiiIc}) is automatically satisfied 
by the fourth vector in Eq.~(\ref{NBases}). Eq.~(\ref{CiiRecEq}) determines 
$\psi^{(4)}_{s+1}$, $\cdots$, $\psi^{(4)}_{p-1}$ recursively 
since $\det(s+n+1-M)\ne0$ for $n=0,1,\cdots,p-s-2$. 
Eq.~(\ref{CiiRecEq}) for $n=p-s-1$ needs some care 
because of $\det(p-M)=0$. It gives $a^{(4,i)}$ without 
ambiguity, whereas $\psi^{(4)}_p$ cannot be determined 
uniquely since we can freely modify $\psi^{(4)}_p$ by 
$\sum_{i=1,2,3}c_i\psi^{(i)}_p$, where $c_i$ are arbitrary 
constants. It can be seen that such an 
arbitrariness does not give any new independent solutions and 
can be removed simply by assuming 
$\psi^{(4)}_p\propto\psi^{(4)}_s$, i.e., the fourth in Eq. (\ref{NBases}). 
Once $\psi^{(4)}_p$ 
and $a^{(4,i)}$ are obtained, we can find $\psi^{(4)}_{s+n+1}$ 
for $n\geq p$. 

\vskip .25cm
We shall be brief since we can infer the solutions for 
any given $M$ in a manner similar to these.

\subsubsection{$n_1=n_2>n_3=n_4$ }
\label{s:Three}

We define $p=n_1$ and $s=n_3$. 
Two solutions $\psi^{(i)}$ ($i=1,2$) with $\lambda=p$ 
can be obtained as in the case of App. \ref{s:One}. Other two solutions 
$\psi^{(i)}$ ($i=3,4$) with $\lambda=s$ 
can be written as 
\begin{alignat}1
  \psi^{(i)}=\sum_{n=0}^\infty\psi^{(i)}_{n+s}r^{n+s}
  +\sum_{j=1,2}a^{(i,j)}\psi^{(j)}\ln r.
\end{alignat}
We can find all the coefficients $\psi^{(i)}_{n+s}$ 
($n>0$) and $a^{(i,j)}$ as in the case of App. \ref{s:Two}.

\subsubsection{$n_1=n_2>n_3>n_4$ }
\label{s:Four}


Defining $p=n_1$ and $s=n_3$, we obtain three 
solutions $\psi^{(i)}$ ($i=1,2,3$) as in the case of App. \ref{s:Three}. 
The fourth solution takes the form 
\begin{alignat}1
  \psi^{(4)}=&\sum_{n=0}^\infty\psi^{(4)}_{n+t}r^{n+t}
  +\sum_{i=1,2}\left(a^{(4,i)}\ln r
    +\frac{1}{2}b^{(4,i)}(\ln r)^2\right)\psi^{(i)} \nonumber \\
  &+a^{(4,3)}\psi^{(3)}\ln r,
\end{alignat}
where $t=n_4$. 
It should be stressed that $(\ln r)^2$ term appears. 
In cases where $M$ has more than two 
different diagonal elements, there appear higher $\ln r$ corrections.

\subsubsection{$n_1>n_2=n_3=n_4$ }
\label{s:Five}

This case is similar to App. \ref{s:Two} or \ref{s:Three}.   
We can find solutions of the form 
\begin{alignat}1
  &\psi^{(1)}=\sum_{n=0}^\infty \psi^{(1)}_{n+p}r^{n+p}, 
\nonumber\\
  &\psi^{(i)}=\sum_{n=0}^\infty\psi^{(i)}_{n+s}r^{n+s}
  +a^{(i,1)}\psi^{(1)}\ln r, \quad (i=2,3,4)
\end{alignat}
where $p=n_1$ and $s=n_2$. 

\subsubsection{$n_1>n_2=n_3>n_4$ }
\label{s:Six}

The solutions can be found in the form 
\begin{alignat}1
  &\psi^{(1)}=\sum_{n=0}^\infty \psi^{(1)}_{n+p}r^{n+p}, 
\nonumber\\
  &\psi^{(i)}=\sum_{n=0}^\infty\psi^{(i)}_{n+s}r^{n+s}
  +a^{(i,1)}\psi^{(1)}\ln r, \quad (i=2,3) 
\nonumber\\
  &\psi^{(4)}=\sum_{n=0}^\infty \psi^{(4)}_{n+t}r^{n+t}
  +\left(a^{(4,1)}\ln r+\frac{1}{2}b^{(4,1)}
    (\ln r)^2\right)\psi^{(1)} 
\nonumber \\
  &\hskip 1cm +\sum_{i=2,3}a^{(4,i)}\psi^{(i)}\ln r,
\end{alignat}
where $p=n_1$, $s=n_2$ and $t=n_4$. 

\subsubsection{$n_1>n_2>n_3=n_4$ }

The solutions can be written as 
\begin{alignat}1
  \psi^{(1)}=&\sum_{n=0}^\infty \psi^{(1)}_{n+p}r^{n+p}, 
\nonumber\\
  \psi^{(2)}=&\sum_{n=0}^\infty\psi^{(2)}_{n+s}r^{n+s}
  +a^{(2,1)}\psi^{(1)}\ln r, 
\nonumber\\
  \psi^{(i)}=&\sum_{n=0}^\infty \psi^{(i)}_{n+t}r^{n+t}
  +\left(a^{(i,1)}\ln r+\frac{1}{2}b^{(i,1)}
    (\ln r)^2\right)\psi^{(1)} \nonumber \\
  &+a^{(i,2)}\psi^{(2)}\ln r, \quad (i=3,4)  
\end{alignat}
where $p=n_1$, $s=n_2$ and $t=n_3$. 

\subsubsection{$n_1>n_2>n_3>n_4$ }

%

\begin{alignat}1
  \psi^{(1)}&=\sum_{n=0}^\infty \psi^{(1)}_{n+p}r^{n+p},
\nonumber \\
  \psi^{(2)}&=\sum_{n=0}^\infty\psi^{(2)}_{n+s}r^{n+s}
  +a^{(2,1)}\psi^{(1)}\ln r, 
\nonumber\\
  \psi^{(3)}&=\sum_{n=0}^\infty \psi^{(3)}_{n+t}r^{n+t}
  +\left(a^{(3,1)}\ln r+\frac{1}{2}b^{(3,1)}
    (\ln r)^2\right)\psi^{(1)} \nonumber \\
  &+a^{(3,2)}\psi^{(2)}\ln r, 
\nonumber\\
  \psi^{(4)}&=\sum_{n=0}^\infty \psi^{(4)}_{n+k}r^{n+k} \nonumber \\
  &+\left(a^{(4,1)}\ln r+\frac{1}{2}b^{(4,1)}
    (\ln r)^2
  +\frac{1}{3}c^{(4,1)}(\ln r)^3\right)\psi^{(1)} \nonumber\\
  &+\left(a^{(4,2)}\ln r+\frac{1}{2}b^{(4,2)}
    (\ln r)^2\right)\psi^{(2)}
  +a^{(4,3)}\psi^{(3)}\ln r, 
\end{alignat}
where $p=n_1$, $s=n_2$, $t=n_3$ and $k=n_4$. 

In any cases, $(\ln r)^n$ terms appear in nonleading order 
of $r$. We thus conclude that the solutions behave as
\begin{alignat}1
  \psi^{(i)}(r)\mathop{\sim}_{r\rightarrow0}
  \psi_{n_i}^{(i)}r^{n_i}, \quad
  (i=1,\cdots,4) .
\end{alignat}
This justifies the counting rule 
of the number of regular solutions at the origin
in Sec. \ref{s:AnaInd}.

\end{document}